\documentclass[12pt]{article}

\usepackage{epsfig} 
\usepackage{amsbsy} 

\def\beq{\begin{equation}}
\def\eeq{\end{equation}}

\def\beqn{\begin{eqnarray}}
\def\eeqn{\end{eqnarray}}

\def\lq{\left[}
\def\rq{\right]}

\def\({\left(}
\def\){\right)}

\def\Dx{\int {\cal D}x }
\def\I{{\cal I}}
\def\Q{{\cal Q} }
\def\P{{\cal P} }
\def\X{{\cal X} }
\def\Y{{\cal Y} }

\newcommand{\n}[1]{\nu_{#1}}

\def\pp#1{\nu_{#1}(\nu_{#1}+1){\bf #1^{\boldsymbol{++}}}}
\def\pps#1{{\bf #1^{\boldsymbol{++}}}}
\def\p#1{\nu_{#1}{\bf #1^{\boldsymbol{+}}}}
\def\ps#1{{\bf #1^{\boldsymbol{+}}}}
\def\m#1{{\bf #1^{\boldsymbol{-}}}}

\def\xbox{{\rm Xbox}}
\def\xtri{{\rm Xtri}}
\def\rtri{{\rm Rtri}}
\def\xbm#1#2#3#4{\xbox_{#1}^{#2}\(#3,#4\)}
\def\dbm#1#2#3{{\rm Dbox}^{#1}\(#2,#3\)}
\def\ssm#1#2{{\rm Sset}^{#1}\(#2\)}
\def\bbm#1#2#3{{\rm Bbox}^{#1}\(#2,#3\)}
\def\xtm#1#2{{\rm Xtri}^{#1}\(#2\)}
\def\tm1#1#2{{\rm Btri}^{#1}\(#2\)}

\def\Xbox{\xbox^D\(\nu_1,\nu_2,\nu_3,\nu_4,\nu_5,\nu_6,\nu_7;s,t\)}

\def\IBP{{\rm integration-by-parts}}

\def\dplus{{\bf d^{\boldsymbol{+}}}} 
\def\dminus{{\bf d^{\boldsymbol{-}}}} 

\def\f21{{_2F_1}}

\def\s{\sigma}

\def\so3#1{\,{\rm S}_{1,\,3}\left(#1 \right)}
\def\st2#1{\,{\rm S}_{2,\,2}\left(#1 \right)}

\newcommand{\Li}[2]{{\mbox{Li}}_{#1}\left(#2\right)}
\newcommand{\snp}[2]{{\mbox{S}}_{#1}\left(#2\right)}
\def\mtos{-\frac{t}{s}}

\newskip\humongous \humongous=0pt plus 1000pt minus 1000pt

\newif\ifdtup

\catcode`\@=11

\@addtoreset{equation}{section}

\def\theequation{\thesection.\arabic{equation}}

\def\@normalsize{\@setsize\normalsize{15pt}\xiipt\@xiipt
\abovedisplayskip 14pt plus3pt minus3pt%
\belowdisplayskip \abovedisplayskip
\abovedisplayshortskip \z@ plus3pt%
\belowdisplayshortskip 7pt plus3.5pt minus0pt}

\def\small{\@setsize\small{13.6pt}\xipt\@xipt
\abovedisplayskip 13pt plus3pt minus3pt%
\belowdisplayskip \abovedisplayskip
\abovedisplayshortskip \z@ plus3pt%
\belowdisplayshortskip 7pt plus3.5pt minus0pt
\def\@listi{\parsep 4.5pt plus 2pt minus 1pt
     \itemsep \parsep
     \topsep 9pt plus 3pt minus 3pt}}

\@twosidetrue

\catcode`@=12

\catcode`\@=11

\def\section{\@startsection{section}{1}{\z@}{3.5ex plus 1ex minus
   .2ex}{2.3ex plus .2ex}{\large\bf}}

\def\thesection{\arabic{section}}
\def\thesubsection{\arabic{section}.\arabic{subsection}}
\def\thesubsubsection{\arabic{section}.\arabic{subsection}.\arabic{subsubsection}}

\def\appendix{\setcounter{section}{0}
 \def\thesection{\Alph{section}}
 \def\theequation{\Alph{section}.\arabic{equation}}
 \def\thesubsection{\Alph{section}.\arabic{subsection}}
\def\thesubsubsection{\Alph{section}.\arabic{subsection}.\arabic{subsubsection}}

\def\section{\@startsection{section}{1}{\z@}{3.5ex plus 1ex minus
   .2ex}{2.3ex plus .2ex}{\large\bf}}
}

\newcommand{\ccaption}[2]{
  \begin{center}
    \parbox{0.85\textwidth}{
      \caption[#1]{\small\it {#2}}}
  \end{center}    }

\def \ep{\epsilon}
\def \to   {\mbox{$\rightarrow$}}

\newcommand\hepph[1]{{\tt hep-ph/#1}}
\newcommand\hepth[1]{{\tt hep-th/#1}}

\def\ord#1{{\cal O}\(#1\)}

\newcount\minutes
\newcount\scratch

\def\timestamp{%
\scratch=\time
\divide\scratch by 60
\edef\hours{\the\scratch}
\multiply\scratch by 60
\minutes=\time
\advance\minutes by -\scratch
---$\,$\hours:\null
\ifnum\minutes< 10 0\fi
\the\minutes}

\begin{document}
\begin{titlepage}
\nopagebreak
{\flushright{
        \begin{minipage}{5cm}
         Freiburg-THEP 00/3\\
         DTP/00/16 \\
         TTP00-07\\
        {\tt hep-ph/0003261}\hfill \\
        \end{minipage}        }

}
\vfill
\begin{center}
{\Large \bf \sc
 \baselineskip 0.9cm
The tensor reduction and master integrals of the two-loop massless crossed
box with light-like legs
          
}
\vskip 0.5cm 
{\large  C.~Anastasiou$^{(a)}$\footnote{e-mail: {\tt
Ch.Anastasiou@durham.ac.uk}}, 
T.~Gehrmann$^{(b)}$\footnote{e-mail: {\tt
Thomas.Gehrmann@phys.uni-karlsruhe.de}}, 
C.~Oleari$^{(a)}$\footnote{e-mail: {\tt Carlo.Oleari@durham.ac.uk}},\\[4mm]
E.~Remiddi$^{(c)}$\footnote{e-mail: {\tt Ettore.Remiddi@bo.infn.it}} and
J.~B.~Tausk$^{(d)}$\footnote{e-mail: {\tt tausk@physik.uni-freiburg.de}}}  
\vskip .2cm 
{$^{(a)}$ {\it Department of Physics, University of Durham, Durham DH1 3LE,
England }}\\ 
{$^{(b)}$ {\it  Institut f\"ur Theoretische Teilchenphysik,
Univ. Karlsruhe, D-76128 Karlsruhe, Germany}}\\
{$^{(c)}$ {\it Dip. di Fisica,
    Universit\`{a} di Bologna and INFN, Sez. di Bologna, I-40126 Bologna,
Italy}}\\  
{$^{(d)}$ {\it Fakult{\"a}t f{\"u}r Physik,
Albert-Ludwigs-Universit{\"a}t Freiburg,
D-79104 Freiburg, Germany }}\\

\vskip
1.3cm    
\end{center}

\nopagebreak
\begin{abstract}
The class of the two-loop massless crossed boxes, with light-like external
legs, is the final unresolved issue in the program of computing the
scattering amplitudes of $2\,\to\, 2$ massless particles at
next-to-next-to-leading order.

In this paper, we describe an algorithm for the tensor reduction of 
such diagrams.
After connecting tensor
integrals to scalar ones with arbitrary powers of propagators in higher
dimensions, we derive recurrence relations from \IBP\ and Lorentz-invariance
identities, that allow us to write the scalar integrals as a combination of two
master crossed boxes plus simpler-topology diagrams.

We derive the system of differential equations that the two master integrals
satisfy using two different methods, and we use one of these equations to
express the second master integral as a function of the first one, already
known in the literature.  We then give the analytic expansion of the second
master integral as a function of $\ep=(4-D)/2$, where $D$ is the space-time
dimension, up to order $\ord{\ep^0}$.
\end{abstract}
\vfill
\vfill
\end{titlepage}
\newpage                                                                     
\section{Introduction}
The increasing precision of high-energy scattering experiments has reached
the level that next-to-next-to-leading theoretical results for $2\,\to\, 2$
scattering processes of massless particles are demanded~\cite{BDK}.
Several steps have already been taken towards this goal: in
Refs.~\cite{Smirnov,Smirnov_Veretin} the tensor reduction of two-loop planar
boxes and the computation of the relevant master integrals have been
given. Reference~\cite{pentabox}  dealt with the tensor reduction of the
pentabox, while other simpler two-loop diagrams obtained by one-loop
insertion into one-loop four-point functions have been treated in
Ref.~\cite{AGO2}. 
 
The final unresolved issue is the tensor reduction of two-loop crossed boxes
of Fig.~\ref{fig:Xboxfig} and the evaluation of the corresponding master
integrals.  One of the master integrals, corresponding to the scalar integral
with all powers of propagators equal to unity, was computed in
Ref.~\cite{Bas} as an analytic expansion in $\ep=(4-D)/2$, where $D$ is the
space-time dimension.

A first result of this paper is to present a method to reduce tensor
integrals of crossed two-loop boxes to two crossed-box master integrals,
plus master integrals for simpler-topology diagrams. We choose as first
master integral the already evaluated~\cite{Bas} crossed box 
with all powers of propagators equal to one, and as second
master integral the scalar integral where the power of the second propagator
is equal to two, all the others being one.

As it is well known~\cite{pentabox,Tar}, tensor integrals can  be
related to scalar integrals with higher powers of the propagators in higher
dimensions. We make use of recurrence relations obtained by \IBP~\cite{ibyp}
and Lorentz-invariance identities~\cite{GR} to connect integrals with
different powers of the propagators and to derive a reduction algorithm that
expresses the generic scalar integral as a function of the two master crossed
boxes, plus simpler sub-topologies.  We explicitly give the equations that
connect the two master integrals in different dimensions (dimensional shift).

Quite in general, it has been shown that the master integrals for any Feynman
graph topology satisfy a system of first-order differential equations on any
of the Mandelstam variables on which they depend~\cite{ER}.  
We derive a system of  two coupled differential equations for the
crossed-box master integrals by two independent methods: first using the
raising and lowering operators and second by taking the on-shell limit of the
differential equations for the crossed box with one off-shell leg~\cite{GR}.

Inserting the analytic expression of the first master crossed box into the
differential equation for the first master integral, we obtain an algebraic
equation for the second master crossed box, that can be solved to give the
analytic expansion in $ \ep$ of the second master integral.  The differential
equation for the second master integral provides an overall check of the
calculation.

Our paper is organized as follows: in Section~\ref{sec:notation} we introduce
the notation we are going to use. In Section~\ref{sec:tensor_redux} we
briefly review how tensor integrals are related to scalar ones, and we give
the reduction algorithm to express these integrals as a combination of the
two master crossed boxes, plus simpler master integrals for topologies with
fewer propagators.  In Sections~\ref{sec:system} and~\ref{sec:diff_eqs} we
derive the differential equations that the two master integrals satisfy in
arbitrary dimensions $D$, using the two different methods.  The analytic
expansion of the second master integral is given in
Section~\ref{sec:an_expans}. For completeness, in Appendix~\ref{app:Xtri} we
present the algorithm for the tensor reduction and for the dimensional shift
of the crossed triangle, since this is one of the non-trivial sub-topologies
produced by the reduction procedure.  Finally we conclude with
Section~\ref{sec:conclusions}.

\section{Notation}
\label{sec:notation}
\begin{figure}[htb]
\centerline{\epsfig{figure=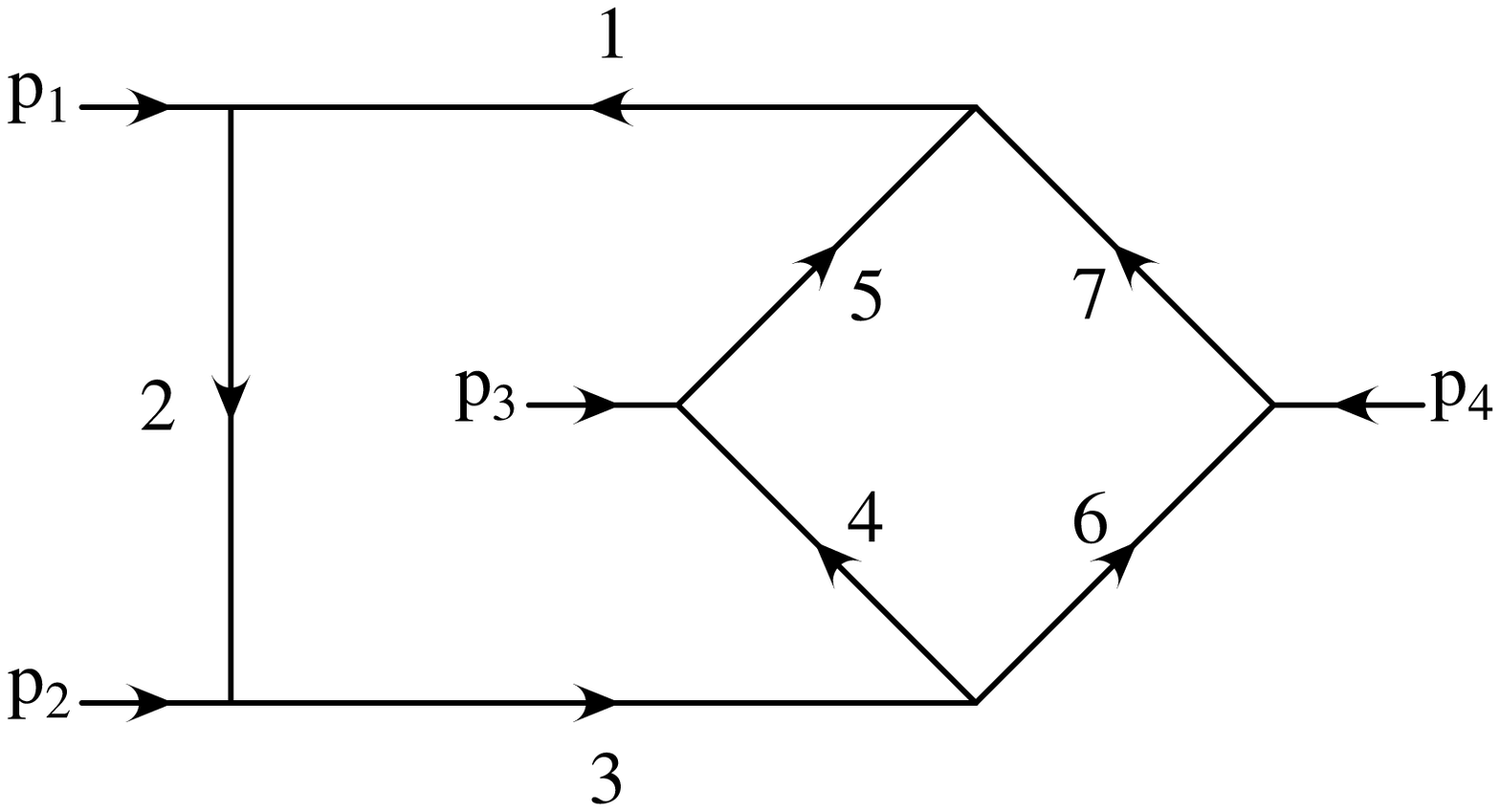,width=0.65\textwidth,clip=}}
\ccaption{}
{ \label{fig:Xboxfig} The generic two-loop crossed  box. }
\end{figure}
We denote the generic two-loop tensor crossed (or non-planar) four-point
function in $D$ dimensions of Fig.~\ref{fig:Xboxfig} with seven propagators
$A_i$ raised to arbitrary powers $\n{i}$ as
\beqn
\label{eq:Xbox_def}
&&\Xbox\lq 1;\, k^\mu;\, l^\mu;\, k^\mu k^\nu;\,k^\mu l^\nu;\,\ldots \rq
\hspace{4cm}
\nonumber\\
&&\hspace{5cm} {}
=\int \frac{d^D k}{i \pi^{D/2}}
\int \frac{d^D l}{i \pi^{D/2}}
~
\frac{\lq 1;\, k^\mu;\, l^\mu;\, k^\mu k^\nu;\,k^\mu l^\nu;\,\ldots \rq}{
A_1^{\nu_1}
A_2^{\nu_2}
A_3^{\nu_3}
A_4^{\nu_4}
A_5^{\nu_5}
A_6^{\nu_6}
A_7^{\nu_7}},
\eeqn
where the propagators are 
\begin{eqnarray}
\label{eq:propagators}
A_1 &=& (k+l+p_{34})^2 + i0, \nonumber \\
A_2 &=& (k+l+p_{134})^2 + i0, \nonumber \\
A_3 &=& (k+l)^2 + i0, \nonumber \\
A_4 &=& l^2 + i0, \\
A_5 &=& (l+p_3)^2 + i0, \nonumber \\
A_6 &=& k^2 + i0, \nonumber \\
A_7 &=& (k+p_4)^2 + i0. \nonumber
\end{eqnarray}
The external momenta $p_j$ are in-going and light-like, $p_j^2=0$,
$j=1\ldots 4$, so that the only momentum scales are the usual
Mandelstam variables $s = (p_1+p_2)^2$ and $t=(p_2+p_3)^2$, together with
$u=-s-t$.  For ease of notation, we define $p_{ij}=p_i+p_j$ and
$p_{ijk}=p_i+p_j+p_k$.  
In the square brackets we keep trace of the tensor
structure that may be  present in the numerator of
Eq.~(\ref{eq:Xbox_def}). When the numerator is unity we have the
scalar integral 
\beq
\Xbox[1] \equiv \Xbox.
\eeq

\section{Tensor reduction using raising and lowering operators}
\label{sec:tensor_redux}
Tensor integrals can be related to combinations of scalar integrals with
higher powers of propagators and/or different values of
$D$~\cite{Tar,pentabox}.  In fact, introducing the Schwinger parameters
$x_i$, we can write the integrand of Eq.~(\ref{eq:Xbox_def}) in the form
\begin{equation}
\frac{1}{A_1^{\nu_1}\ldots A_7^{\nu_7}}
=  \Dx~\exp\left ( \sum_{i=1}^7 x_iA_i \right),
\end{equation}
where
\beqn
\label{eq:Dx}
\Dx &=&  \prod_{i=1}^7 \frac{(-1)^{\nu_i}}{\Gamma(\nu_i)}
\int^{\infty}_0 dx_i\, x_i^{\nu_i-1},
\\
\sum_{i=1}^7 x_iA_i &=& a\, k^2 + b\, l^2 + 2\, c \,k \cdot l + 2\, d
\cdot k + 2 \, e \cdot l+ f, 
\label{eq:sumaixi}
\eeqn
and
\begin{eqnarray}
\label{eq:abc}
a &=& x_1+x_2+x_3+x_6+x_7\nonumber \\
b &=& x_1+x_2+x_3+x_5+x_4\nonumber \\
c &=& x_1+x_2+x_3\nonumber \\
d^\mu &=& x_1 p_{34}^\mu + x_2 p_{134}^\mu + x_7 p_4^\mu \nonumber \\
e^\mu &=& x_1 p_{34}^\mu  +x_2 p_{134}^\mu + x_5 p_3^\mu \nonumber \\
f &=& x_1 \,s.
\end{eqnarray}
We can diagonalize the exponent with the change of variables
\begin{eqnarray}
\label{eq:k1shift}
k^\mu &\to& \(K -\frac{c L}{a} +\X \)^{\mu}, \\ 
\label{eq:k2shift}
l^\mu &\to& \(L + \Y\)^{\mu},
\end{eqnarray}
where
\beqn
\label{eq:Xbox_Xmu}
\X^{\mu} &=&  \frac{1}{\P} 
    \Big\{ -\lq (x_2+x_1) (x_7+x_5+x_4)+(x_5+x_4+x_3) x_7 \rq p_4^{\mu}
    + \lq x_3 x_5-x_4(x_2+x_1) \rq p_3^{\mu} 
\nonumber\\
  &&\phantom{\frac{1}{\P} \Big\{  } 
   - x_2 (x_5+x_4) p_1^{\mu} \Big\},
\\
\Y^{\mu} &=& \frac{1}{\P} \Big\{
        \lq x_3 x_7-x_6(x_2+x_1) \rq p_4^{\mu}
	-\lq (x_2+x_1) (x_7+x_6+x_5)+(x_7+x_6+x_3) x_5) \rq p_3^{\mu}
\nonumber\\
  &&\phantom{\frac{1}{\P} \Big\{  } 
	-x_2 (x_7+x_6)p_1^{\mu} \Big\},
\eeqn
and
\beq
\label{eq:P}
\P = \(x_7+x_6+x_5+x_4\) \(x_3+x_2+x_1\)+\(x_5+x_4\)\(x_7+x_6\).
\eeq
The generic tensor integral (dropping all the dependences on $\n{i}$ and on the
external scales) becomes
\beqn
&&\xbox^D[k^{\mu_1} \ldots
      k^{\mu_m} l^{\nu_1}  \ldots l^{\nu_n} ] 
=\Dx 
\int \frac{d^D K}{i \pi^{D/2}}
\int \frac{d^D L}{i \pi^{D/2}} \nonumber \quad\quad\quad
\\&& \hspace{2.5cm} 
\times \(K -\frac{c L}{a} +\X \)^{\mu_1} \ldots 
\(K -\frac{c L}{a} +\X \)^{\mu_m}
\(L + \Y\)^{\nu_1} \ldots \(L + \Y\)^{\nu_n}
\nonumber
\\&& \hspace{2.5cm} 
\times\exp\( a K^2+\frac{\P}{a}L^2 +\frac{\Q}{\P} \),
\eeqn
where
\beq
\label{eq:Q}
\Q = x_2 (x_5 x_6-x_4 x_7)t + 
    \lq x_1 x_3 (x_7+x_6+x_5+x_4)+x_3 x_5 x_7-x_2 x_4 x_7+x_1 x_4 x_6 \rq s.
\eeq
The integration over the loop momenta $K$ and $L$ is now straightforward. For
example, the scalar integral and  the two one-index tensor integrals are
given by
\beqn
\label{eq:Schwinger}
\Xbox &=& \Dx ~ \I,\\
\Xbox[k^\mu] &=& \Dx ~ \X^\mu \I,\\
\Xbox[l^\mu] &=& \Dx ~ \Y^\mu \I,
\eeqn
where
\beq
\label{eq:def_I}
\I =
\frac{1}{\P^{D/2}}
~\exp\left (\frac{\Q}{\P} \right).
\eeq
Recalling the definition~(\ref{eq:Dx}), we see that we can absorb the factors
$x_i$ of $\X^{\mu}$ and $\Y^{\mu}$ into ${\cal D}x$, increasing the power
of the $i$-th propagator by one
\begin{equation}
\label{eq:times_x_i}
\frac{(-1)^{\nu_i} x_i^{\nu_i-1}}{\Gamma(\nu_i)} ~x_i \ \Longrightarrow \ 
-\nu_i \frac{(-1)^{\nu_i+1} x_i^{\nu_i}}{\Gamma(\nu_i+1)} \equiv 
-\nu_i {\bf  i^{\boldsymbol{+}}},
\end{equation}
where
\beq
{\bf i^{\boldsymbol{\pm}}} \xbox^D\(\ldots, \nu_i, \ldots\) = \xbox^D\(\ldots,
\nu_i\pm1, \ldots\),
\eeq
while the factor $\P$ can be absorbed into $\I$ (see Eq.~(\ref{eq:def_I}))
\begin{equation}
\frac{1}{\P^{D/2}} ~\frac{1}{\P} \ \Longrightarrow\ 
\frac{1}{\P^{(D +2)/2}},
\end{equation}
so that $1/\P$ acts as a dimension increaser
\beq
\label{eq:P_and_dm}
\frac{1}{\P}  \ \Longrightarrow\ \dplus, \quad\quad\quad 
\P  \ \Longrightarrow\ \dminus,
\eeq
where
\beq
\label{eq:pm_d}
  {\bf d^{\boldsymbol{\pm}}} \xbox^D =  
   \xbox^{D\pm 2}.
\eeq
For example, using the expression for $\X^{\mu}$ of Eq.~(\ref{eq:Xbox_Xmu}),
we can write
\beqn
\label{eq:Xbox_kmu}
\xbox^D[k^{\mu}] &=& 
    \Big\{ \lq \p3 \p5-\p4\(\p2+\p1\) \rq p_3^{\mu} 
 - \p2 \(\p5+\p4\) p_1^{\mu}
\nonumber\\
  &&{} -\Big[ \(\p2+\p1\) \(\p7+\p5+\p4\)\nonumber\\
  &&{}+\(\p5+\p4+\p3 \) \p7\Big] p_4^{\mu}
 \Big\} \,\xbox^{D+2}[1].
\eeqn

The task to compute tensor integrals has then been moved to the computation
of scalar integrals with higher powers of the propagators in higher
dimensions.

In the next sections, we will follow the procedure already used in
Ref.~\cite{Smirnov_Veretin}:
\begin{itemize}
\item[1)]
we first reduce the powers of the propagators of the generic scalar diagram
and we  express it in terms of a finite set (basis) of master
diagrams;
\item[2)]
then we relate the master integrals in higher dimensions to the master
integrals in $D$ dimensions (dimensional shift).
\end{itemize}

\subsection{The scalar crossed-box reduction}
\label{sec:reduction}
The strategy to reduce the generic scalar integral to a linear combination of
master ones is based on identities that relate scalar integrals with
different powers of propagators.
Some of these identities can be obtained using the \IBP\
method~\cite{ibyp}.  Since there are three external independent momenta and
two loop momenta, we can build ten identities,  imposing that
\beq
\int \frac{d^D k}{i \pi^{D/2}}
\int \frac{d^D l}{i \pi^{D/2}}
~ \frac{\partial}{\partial a^{\mu}} 
\lq b^{\mu} \,f(k,l,p_i) \rq = 0,
\eeq
where
\beqn
a^{\mu} &=& k^{\mu},\, l^{\mu}\\
b^{\mu} &=& k^{\mu},\,l^{\mu},\, p_1^{\mu},\, p_3^{\mu},\,p_4^{\mu}\\
f(k,l,p_i) &=& \frac{1}{
A_1^{\nu_1}
A_2^{\nu_2}
A_3^{\nu_3}
A_4^{\nu_4}
A_5^{\nu_5}
A_6^{\nu_6}
A_7^{\nu_7}}.
\eeqn
Not all the scalar products that appear in the application of the \IBP\
identities can be written in terms of combinations of propagators: with the
two loop momenta and with the three linearly independent external ones, we
can form nine scalar products involving $k$ and $l$. Since we have seven
linearly independent propagators, we are left with two irreducible scalar
products in the numerator, that we choose to be $\(l\cdot p_1\)$ and
$\(l\cdot p_4\)$.

Three more identities among the scalar integrals can be derived if we exploit
the Lorentz invariance of the Feynman diagrams~\cite{GR}.  In fact, since
the Feynman integral is a function only of scalar products of the external
momenta, it is invariant under the (infinitesimal) rotation
\beq
p_i^{\prime\mu} = \Lambda^{\mu}_{\ \nu} \,p_i^\nu, \quad\quad 
\Lambda_{\mu\nu} = g_{\mu\nu}+\delta\,\ep_{\mu\nu}, \quad\quad 
\ep_{\mu\nu} = -\ep_{\nu\mu}.
\eeq
We can then write
\beq
\label{eq:Lor_invar}
\int \frac{d^D k}{i \pi^{D/2}} \int \frac{d^D l}{i \pi^{D/2}} \ f(k,l,p_i) =
\int \frac{d^D k}{i \pi^{D/2}} \int \frac{d^D l}{i \pi^{D/2}} 
\ f(k,l,p_i^\prime).
\eeq
Expanding in a Taylor series in $\delta$ the right-hand side of
Eq.~(\ref{eq:Lor_invar}), we obtain
\beq
\label{eq:Lor_tensor}
\int \frac{d^D k}{i \pi^{D/2}} \int \frac{d^D l}{i \pi^{D/2}} \, 
\sum_{j=1}^{3}\, \frac{\partial f(k,l,p_i) }{\partial p_j^\mu} \,
\ep^{\mu}_{\ \nu} \,p_j^{\nu} = 0.
\eeq
Using the three independent external momenta, we can build three independent
antisymmetric tensors
\beqn
\ep^{\mu\nu}_1 &=& p_1^\mu\, p_2^\nu - p_2^\mu\, p_1^\nu ,\nonumber\\
\ep^{\mu\nu}_2 &=& p_1^\mu\, p_3^\nu - p_3^\mu\, p_1^\nu ,\nonumber\\
\ep^{\mu\nu}_3 &=& p_2^\mu\, p_3^\nu - p_3^\mu\, p_2^\nu ,\nonumber
\eeqn
that, once inserted into Eq.~(\ref{eq:Lor_tensor}), give rise to three more
identities.

Taking linear combinations of these thirteen equations, we can build the
following system
\beqn
\label{eq:b1}
&& s\p1- \p7\m6 - \p5\m4 -\(\p2+\p1\) \m3 - \n{1257} -2 \n{346} +2 D= 0 \\  
\label{eq:b2}
&& s \p3 -\p6 \m7-\p4 \m5 -\(\p3 + \p2\) \m1 -\n{2346} -2 \n{157} +2 D= 0 \\   
\label{eq:b3} 
&& 2 \(l\cdot p_4\) \p4 \,- \(\p6 +\p5\) \m7+ \p5\m1+\p4\(\m3  - \m6\)
\nonumber\\
&&\hspace{1.8cm}{}  -\n{456} -2 \n{7}+ D = 0 \\  
\label{eq:b4}    
&&2 \( l\cdot p_4\)  \p5 + \p7\m6+ \p5 \(\m7- \m1 + s\) +\p4 \(\m6- \m3\)
\nonumber\\
&&\hspace{1.8cm}{}  +\n{457}+2\n{6} -D = 0 \\  
\label{eq:b5}   
&&2 \(l\cdot p_4\)  \p6  + \p7\(\m5- \m1\) 
  + \p6 \( \m7 + \m5- \m1 + s\) +\p4 \m5\nonumber\\
&&\hspace{1.8cm}{} +\n{47}+2 \n{5} -D = 0 \\ 
\label{eq:b6}   
&& 2 \(l\cdot p_4\) \p7 +\p6\(\m3 - \m4\)  - \p7\( \m6 + \m4-\m3\)-\p5\m4 
\nonumber\\
&&\hspace{1.8cm}{}   - \n{56}-2 \n{4}  +D = 0 \\ 
&& 2s\(l\cdot p_4 \)\p2  +4 s \(l\cdot p_4\)  \p3 
+ \p2 \lq \(t+s\) \(s+ \m1\) - t \m3 \rq  
 \nonumber \\
&& \hspace{1.8cm}{}  + s \( 2 \p6  +2 \p3  + \p2\) \m7 
- s\(2  \p3  +   \p2\) \m6 \nonumber\\
&&\hspace{1.8cm}{} -\(2 D-2 \n{13457}-\n{2}\) s= 0 \\ 
&&2  s \( l\cdot p_4\) \p3 -2  s \(l \cdot p_4\) \p1
+ (t+s)\p2 \(\m1 - \m3\) +s \( \p1 - \p3\) \m6 \nonumber\\
&&\hspace{1.8cm}{} + s \(\p6  + \p3  - \p1 \)\m7 
 -s \p5 \m4 -s \p1 \m3  \nonumber\\
&&\hspace{1.8cm}{} 
   -\(\n{6}-\n{15}\) s  = 0 \\     
&&  2 \(l \cdot p_4\) \p3 -2 \(l\cdot p_1\)\p2 + \(t+s\) \p2 
+\(\p6 +\p3 +\p2 +\p1\) \m7 \nonumber \\
&&\hspace{1.8cm}{}  -\(\p2 +\p1\) \m5-\p3 \m4 +\n{1236} +2 \n{7}-D  = 0 \\    
&&2 s \(l\cdot p_4\) \p3 -2 s\(l\cdot p_1 \) \p7
+ \p7 \lq \(t+s\) \(s  - \m1\) +t \m5 +  s\m2 \rq \nonumber \\
&&\hspace{1.8cm}{}  + \p6 \( t \m4 -  t\m3  +s \m7 \)
+s\p3 \( \m7 - \m6\) +(t+s) \p2\m1 +s\p1 \m2  \nonumber\\
&&\hspace{1.8cm}{} + s \( \n{126} + 2 \n{3457} - 2 D \) 
    - t\( D   -\n{67} +  \n{2} -2 \n{45}\) = 0 \\     
&& 2  s  \(l\cdot p_4\) \p3 -2s\( l\cdot p_1 \)\p5 
+ \(t+s\)\p5 \( \m7 -\m1+s\)  + (t+s)\p2 \m1 \nonumber \\
&&\hspace{1.8cm}{}
+(t+s)\p4 \(\m6  -  \m3\)  +s \(\p6  + \p3\) \m7   -s\p3 \m6 
 -s\p1 \m2 \nonumber\\
&&\hspace{1.8cm}{} + s \(\n{1456}- \n{2} +2  \n{7} -D \)  
 + t \(2  \n{67}  +  \n{45} - \n{2} -D \) = 0 \\    
&& 2 s \(l\cdot p_4\) \p3 +2 s\( l\cdot p_1\)\p6  
+ \p6 \lq t \(\m4-  \m3\) -  s\( \m2-  \m1 +s\) \rq- s\p4 \m5 \nonumber\\
&&\hspace{1.8cm}{} 
+ s\p3\( \m7 -\m6  -  \m2 \)+t \p7 \( \m5-  \m1  \)
 +(t+s)\p2\m1 \nonumber\\
&&\hspace{1.8cm}{} 
  + s \(\n{34} -\n{2}\) + t \( \n{67} +2  \n{45}-\n{2} - D \)  = 0 \\ 
\label{eq:b13}
&&2 s \(l\cdot p_4\) \p3 + 2 s \(l\cdot p_1\) \p4
+\p4 \lq (t+s) \(\m6-  \m3\) +s \m5\rq  +2 s\p6 \m7 \nonumber\\
&&\hspace{1.8cm}{} 
+(t+s)\p5 \(\m7  - \m1 \)
+ (t+s)\p2\m1 
+ s \p3 \( \m7- \m6  + \m2\) \nonumber\\
&&\hspace{1.8cm}{} 
  + s \( \n{23} +2 \n{146} +3  \n{5}+4 \n{7} - 3 D \) 
  +t \( \n{45}-\n{2} + 2 \n{67} - D \) = 0, 
\eeqn
where we have introduced the shorthand $\n{ij}=\n{i}+\n{j}$, $\n{ijk} =
\n{i}+\n{j}+\n{k}$, etc.
Each equation acts on the integrand of the generic crossed box  before
any loop integration has taken place.

Equations~(\ref{eq:b1}) and~(\ref{eq:b2}) of the system, being independent
of the two irreducible scalar products, need no further
manipulation, and can be rewritten in the form
\beqn
\label{eq:p1}
 s\p1 &=&  \p7\m6 + \p5\m4 +\(\p2+\p1\) \m3 + \n{1257} +2 \n{346} -2 D, \\  
 s\p3 &=&  \p6\m7 + \p4 \m5 +\(\p3+\p2\) \m1 +\n{2346} +2 \n{157} -2 D.
\eeqn
By repeated application of these two identities, we can reduce $\n{1}$ and
$\n{3}$ to one. During this process, the generic scalar box  is
expressed as a linear combination of crossed-box diagrams with
$\n{1}=\n{3}=1$ and diagrams belonging to simpler topologies, that originate
when powers of propagators are reduced (pinched) to zero by the
decreasing operators.  We will deal with the pinched diagrams later,
concentrating now on the reduction of the remaining propagators.
 
In order to use the other equations of the system,
we have to eliminate the irreducible scalar products in the numerator.

For example, applying the operator $\p{7}$ to Eq.~(\ref{eq:b3}) and $\p{4}$
to Eq.~(\ref{eq:b6}), and taking the difference, we get 
\beqn
\label{eq:4567_a}
&& \(D-2 \n{7}-\n{56}-2\) \p7 - \(D-\n{56}-2 \n{4}-2\)\p4	
 - \(\n{7}-\n{4}\)\(\p5+\p6\) \nonumber \\
&&\hspace{1.8cm}{} 
 +\p5 \p7\m1 - \p4 \p6\m3=0.
\eeqn
In the same way, we can apply $\p{6}$ to Eq.~(\ref{eq:b4}) and $\p{5}$ to
Eq.~(\ref{eq:b5}) and take the difference, to obtain
\beqn
\label{eq:4567_b}
&&\(D-\n{47}-2 \n{5}-2\)\p5  -  \(D-\n{47}-2 \n{6}-2\)\p6
 +\(\n{6}-\n{5}\) \(\p4+\p7\)\nonumber \\
&&\hspace{1.8cm}{} 
  + \p5 \p7\m1 - \p4 \p6\m3=0.					 
\eeqn
Combining Eq.~(\ref{eq:4567_a}) and~(\ref{eq:4567_b}) to eliminate $\p5$,
we have
\beqn
\label{eq:p4}
\p4
 &=& \frac{\(D-2\n{57}-2\)}{\(D-2\n{45}-2\)}\p7 
-2\frac{\(\n{7}-\n{4}\)}{\(D-2\n{45}-2\)}\p6 \nonumber\\
&& {}
+\frac{1}{\(D-\n{4567}-2\)} \( \p5\p7\m1-\p4\p6\m3 \),
\eeqn
that can be used to reduce $\n{4}$ to one, at the expense of increasing
$\n{6}$ and $\n{7}$.  If, on the other hand, we eliminate $\p4$, we obtain
the symmetric equation that can reduce $\n{5}$ to one.  At this point, all
the powers of the propagators except $\n{2}$, $\n{6}$ and $\n{7}$ have been
reduced to one.

In the same spirit  we can derive
\beqn
\label{eq:reduce26}
 s t \p2 \p6 &=&   \(2\n{1567} +\n{2}+2-2 D\) s  \p6
         + \(\n{467} +2 \n{5}-D\) s\p2 \nonumber\\
&& {}
    -2\(D-\n{467} -2 \n{5}\)  
    \( \p6 \m7 + \p4 \m5 + \p3 \m1  + \p2 \m1  \)\nonumber\\
&&{}
   +s \( 2 \p3+ \p2 \) \lq\(\p7  + \p6 + \p4 \) \m5 
   - \p7 \m1  \rq\nonumber\\
&& {}					
  + t \p2 \p6 \(  \m3 - \m1 \)+2 s \p4 \p6 \m5	\nonumber\\
&&{}								
  +2 \(D-\n{467}-2 \n{5}\) \(2 D-2\n{157}-\n{2346}\),
\eeqn
that, together with  the symmetric one for $\p2\p7$  and with
\beqn
\label{eq:reduce67}
s \p6 \p7  &=& \(D-\n{567}-2 \n{4}-1\) \p6 
   +\(D-\n{467}-2 \n{5}-1\) \p7 \nonumber\\
&&{}
    + \p6 \p7 \( \m3 +\m1   -\m5 -\m4 \)-\p4\p7 \m5  -\p5 \p6 \m4	    
\nonumber\\
&&{}
+ \pp7\(\m1- \m5 \) + \pp6 \( \m3 -\m4\) ,
\eeqn
reduces all powers except one ($\n{2}$ or $\n{6}$ or $\n{7}$) to unity.

We can decrease $\n{2}$ at the expense of increasing $\n{6}$ and
$\n{7}$ using
\beqn
\label{eq:reducep2}
&& \lq\(\n{4}-\n{7} +2\n{23}+2-D\) s + \(\n{45} - \n{67}\)  t \rq
\p2 
= \(\n{5}-\n{3}\) s \p4  +\(\n{7}-\n{3}\) s \p6
\nonumber\\
&&\hspace{1.5cm}{}
 -\(D-2 \n{7}-\n{16}-2\) s\p7  - \(D-2\n{5}-\n{14}-2\) s\p5 
\nonumber\\
&&\hspace{1.5cm}{}
+(t+s) \p2\lq \p7 \( \m1  - \m5 \) + \p4 \(\m6-\m3\) \rq
\nonumber\\
&&\hspace{1.5cm}{}
+t \p2 \lq \p5 \( \m7 -\m1 \)+ \p6\( \m3 -   \m4 \)   \rq
\nonumber\\
&&\hspace{1.5cm}{}
 - s \p1\(  \p7 \m5   +   \p5 \m7 \)
\nonumber\\
&&\hspace{1.5cm}{}
+ s \p3 \lq  \(2\p7 + \p4 \) \m6  + \( 2 \p5+\p6 \) \m4 \rq
\nonumber\\
&&\hspace{1.5cm}{}
 -\(2 D-2\n{57}-3\n{46}\) \lq \p6 \m7 +\p4 \m5+ \(\p3  +\p2\) \m1 \rq
\nonumber\\
&&\hspace{1.5cm}{}
 + \(\n{57}-2 \n{2}\) \lq  \p7 \m6+\p5\m4  + \(\p2  +\p1\) \m3 \rq
\nonumber\\
&&\hspace{1.5cm}{}
+4 D^2-2 \(5 \n{57}+4 \n{46} + \n{3}-\n{2}+2 \n{1}\) D 
+ \n{7} \( 5 \n{17} +10 \n{456}  +4 \n{3}+\n{2}  \)
\nonumber\\
&&\hspace{1.5cm}{}
+ \n{6} \( 3\n{6} + 10 \n{5} + 6 \n{14} +3 \n{3} -\n{2} \) 
+ \n{5} \( 5\n{5} +10 \n{4} +4 \n{3} +\n{2} +5 \n{1} \)
\nonumber\\
&&\hspace{1.5cm}{}
+ \n{4} \( 3\n{4} +3 \n{3} -\n{2} +6 \n{1} \)
-2 \n{2} \(  2\n{3}+ \n{12}  \).
\eeqn


The power of the seventh propagator can be reduced with
\beqn
\label{eq:7pp}
&& s (t+s) \pp7 = \s \p7 -(\n{7}-1) \frac{\( D -6 \) t+ 
  \(5 D -2 \n{7} -26\)s}{D-2 \n{7}-6}\ps6 
\nonumber\\
&&{}\hspace{1.5cm}{}  
	+\rho 
 \frac{\( 2 D -\n{7} -11 \)t + \(3 D -2 \n{7} -15 \)s}{D-\n{7}-5}\ps5 \p7  \m1 
\nonumber\\
&&{}\hspace{1.5cm}{}  
 + \rho \frac{ \( D -2 \n{7} -4 \)t+ \(5 - D \)s}{D-\n{7}-5} \ps4 \ps6 \m3
	+ \rho (t+s) \Bigl\{
-\pp7\m6  	
\nonumber\\
&&{}\hspace{1.5cm}{}  
  + \ps5 \p7 \( \m3-\m4-\m6\) + \ps4 \p7  \lq 2\(  \m3  - \m6\) + \m1\rq
\nonumber\\
&&{}\hspace{1.5cm}{}  
	-\ps2 \( \p7  + \ps4 \)\m1  + \ps5 \ps6 \m3
	-2  \,\pps5 \m4\Bigr\}
\nonumber\\
&&{}\hspace{1.5cm}{}  
	-2 \n{7}\frac{t+s}{D-2 \n{7}-6} \( \ps4  \p7  \m5
	+\ps5 \ps6\m4  \)+(t+s) \Bigl[ 2 \, \pps6 \( \m4- \m3 \)
\nonumber\\
&&{}\hspace{1.5cm}{}  
	+ \pp7 \(\m5 - \m1\)
	+ \ps6 \p7 \(\m5+ \m4 -  \m3- \m1\)	\Bigr]
\nonumber\\
&&{}\hspace{1.5cm}{}  
	+ \rho s \Bigl[  \ps3 \ps5 \(\m1 -\m4 -\m7\)-\ps3 \ps6 \( \m4 + \m7 \) 
        -\ps3 \p7\( \m4  + \m2\) 
\nonumber\\
&&{}\hspace{1.5cm}{}  
	- \ps4 \( \p7 \m2 + \ps6 \m7 +\ps1 \m2 + \ps3  \m7\)\Bigr] 
\nonumber\\
&&{}\hspace{1.5cm}{}  
	+(2 D-3 \n{7}-7) \rho \Bigl[ \ps6 \m7 + \ps4 \m5 
	+ \( \ps3 + \ps2\) \m1
        -2 (D-\n{7}-4) \Bigr],\quad\quad\quad
\eeqn
where we have introduced the shorthands
\beqn
\rho &=& \frac{D-6}{D-2\n{7}-6}\nonumber\\
\s &=& \frac{ \(5 D^2-8 \n{7} D-50 D+2 \n{7}^2+42 \n{7}+124\) s + 
               \(2 D^2-3 \n{7} D-21 D+18 \n{7}+54\) t}
        {D-2 \n{7}-6}.\nonumber
\eeqn
Equation~(\ref{eq:7pp}) is not as general as the previous ones since we have
set all the powers of the other propagators to unity.  In addition, since
this equation contains $\pps{7}$, we cannot always reduce $\n{7}$ to one, but
we also have integrals where $\n{7}=2$.  A similar identity can be obtained
by symmetry for $\pps6$, so that we are left with three integrals:
$\xbox^D(1,1,1,1,1,1,1;s,t)$, $\xbox^D(1,1,1,1,1,1,2;s,t)$ and
$\xbox^D(1,1,1,1,1,2,1;s,t)$.

The last step is to write the integral with $\n6=2$ as a combination of the
other two. This can be done with the identity that links $\ps6$ with
$\ps7$. We derived such an identity, equating the expressions obtained  by
acting with $\p{7}$ on $\p{2} \p{4}$ and by acting  with $\p2$ on $\p4 \p7$ 
\beqn
&&(D-6) (D-5) \frac{t}{t+s}\ps6 = (D-6) (D-5) \ps7 -4 \frac{(D-5)^3}{t+s}
\nonumber \\
&&\hspace{1cm}{}
        -  \Big( \ps5\ps7+\ps1 \ps7
       +\ps1 \ps5
      +\ps1 \ps4\Big) \m6 - \Big( \ps3 \ps7 +\ps4 \ps6 
       +\ps3 \ps6+\ps3 \ps4 \Big)\m5
\nonumber \\
&&\hspace{1cm}{}	
+ \Big(\ps3 \ps4+\ps4 \ps6
          +\ps5 \ps6 +\ps3 \ps6\Big)\m1
	-(D-7) \ps4 \ps7 \m1
\nonumber \\
&&\hspace{1cm}{}	
          -\frac{1}{2} \Big(\ps2 \ps7 \m6+\ps2 \ps5 \m6 +\ps4 \ps7 \m5
          +\ps2  \ps6\m5-\ps1 \ps4 \m3\Big)
\nonumber \\
&&\hspace{1cm}{}	
        -\frac{t+s}{2} \Big[ \ps2 \ps4  \ps7 \( \m6+\m5\) 
        +2\(\pps2 \ps7 -\ps2 \ps4 \ps7 + \pps2 \ps4\)  \m1\Big]
\nonumber \\
&&\hspace{1cm}{}	
	-\frac{s}{2}\Big\{\ps2 \Big[ \ps3 \( \ps6 + \ps5 
         + \ps4\)\m7 +\ps4 \ps6 \m7
         + \ps3 \(\ps7  +\ps6 + \ps5\) \m4  \Big]
\nonumber \\
&&\hspace{1cm}{}	          
        + \ps3 \ps5 \(2 \, \ps7 - \ps2\) \m1 \Big\}
	-(D-6) \frac{(t+2 s)}{2 (t+s)}\Big[\ps2 \( \ps7\m4+\ps4 \m7+\ps5\m4 \) 
\nonumber \\
&&\hspace{1cm}{}	
        +2\Big(\ps1 \ps7+\ps5 \ps7+\ps1 \ps6+\ps1 \ps5\Big)\m4 
      -2 \,\ps1  \ps6\m3\Big] 	-\frac{2 (D-5) t+s}{2 s} \ps2 \ps7 \m5
\nonumber \\
&&\hspace{1cm}{} 
-(D-6)\Big[ \ps4 \ps7\m3
       + \(\ps3 \ps4+\ps4 \ps6+\ps3 \ps6+\ps3 \ps5\)\m7  \Big]
\nonumber \\
&&\hspace{1cm}{}
       +(D-6) \frac{s}{t+s}\Big[\ps7\ps3 \m4
      +\Big( \ps7\ps3+\ps7\ps4 
        +\ps1 \ps4\Big) \m2 \Big]         +\frac{2 D-13}{2} \ps4  \ps7\m6
\nonumber \\
&&\hspace{1cm}{}	
	+(D-5) \frac{(D-5) t+(2 D-11) s}{s(t+s)}
        \Big[\(\ps2+\ps1\)\m3 
        +\ps7\m6 \Big]
\nonumber \\
&&\hspace{1cm}{}	
	+(D-5) \frac{t}{t+s}\Big[2 \,\ps3 \ps7 \m6+\ps3 \ps4 \m6-\ps1 \ps7 \m5-
        \ps1 \ps5 \m7\Big]	+\frac{t}{2 s} \ps2  \ps5 \m3
\nonumber \\
&&\hspace{1cm}{}	
	-(D-5) \frac{(D -5) t-s}{s(t+s)}\Big[\ps4 \m5+ \ps2\m1+\ps6 \m7+
         \ps3\m1\Big] 	-\frac{2 D-15}{2} \ps3 \ps7 \m1
\nonumber \\
&&\hspace{1cm}{}	
	+(D-5)\Big[\m1 \(\ps3 \ps5+2 \,\pps2+\ps2 \ps3\)\Big]+
	\frac{t}{2}\Big[\ps2 \(\ps5 \ps7\m1 + \ps4 \ps6\m3\)\Big]
\nonumber \\
&&\hspace{1cm}{}
         +\frac{5 t-2 (D-7) s}{2(t+s)} \ps5 \ps7\m1
       	-\frac{(2 D-9) t^2-(D-5) s t-2 (D-5) s^2}{2 s (t+s)}\ps2 \ps5 \m1 
\nonumber \\
&&\hspace{1cm}{}	
	-(D-6) \frac{t+2 s}{2 s}\ps2 \ps7 \m3 
	-\frac{2 (D-5) t^2+(D-6) s t+2 (D-6) s^2}{2 s (t+s)}\ps2\ps6 \m4 
\nonumber \\
&&\hspace{1cm}{}
        + \frac{(3 D-16) t+2 (D-5) s}{2 s} \ps2 \ps7\m1+
	\frac{(D-4) t+2 s}{2(t+s)}\ps2 \ps6 \m7
\nonumber \\
&&\hspace{1cm}{}	
       +\frac{ 2 (D-5) t^2-(D-6) s t-2 (D-6) s^2}{2 s (t+s)}\ps2 \ps5 \m7
	-(4 D-21) \frac{t}{2(t+s)}  \ps4 \ps6 \m3
\nonumber \\
&&\hspace{1cm}{}
        + \frac{(D-5) t+(D-6) s}{t+s} \ps3\ps6 \m4 
	-\frac{ (D-4) t-2 (D-6) s}{2 s} \ps2\ps4 \m3  
\nonumber \\
&&\hspace{1cm}{}	
	+\frac{(2 D-11) t^2+(D-7) s t+2 (D-6) s^2}{2 s (t+s)} \ps2 \ps6\m3 +
	\frac{t+2 s}{2 s} \ps2 \ps6 \m1 
\nonumber \\
&&\hspace{1cm}{}	
        +\frac{2 (D-5) t+(D-6) s}{t+s} \ps3 \ps5 \m4  +
	(D-5) \frac{ (D-5) t+(2 D-11) s}{s (t+s) }\ps5\m4 
\nonumber \\
&&\hspace{1cm}{}
        + \frac{2 D-11}{2} \ps2 \ps4 \m5 
	-\frac{ (D-6) t-2 s}{2 s}  \ps2 \ps4\m1	
        +\frac{2 (D-5) t-s}{2 s} \ps2\ps4 \m6, 
\eeqn
where we have set all the powers of the propagators to unity.

\begin{figure}[htb]
\centerline{\epsfig{figure=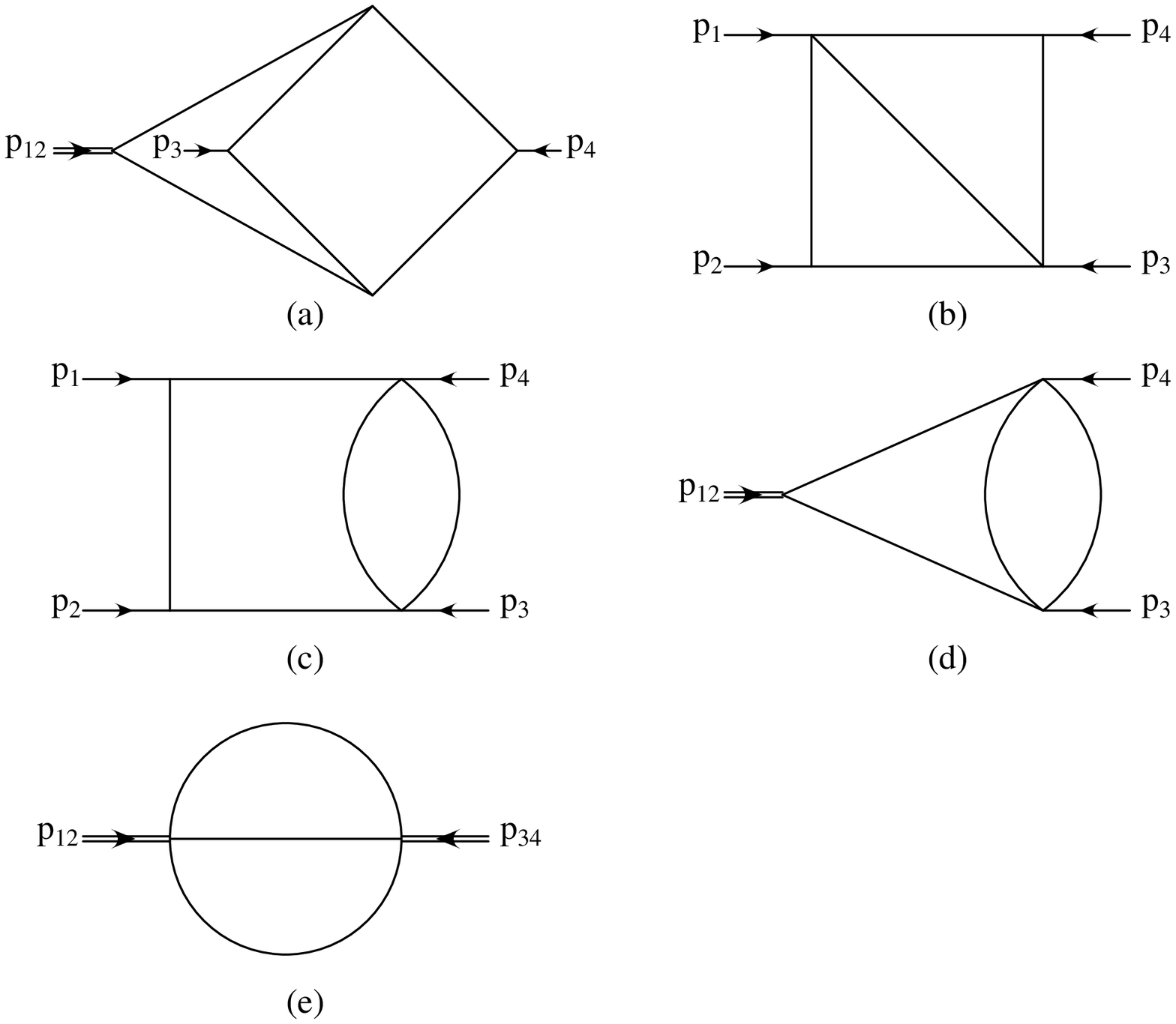,width=0.85\textwidth,clip=}}
\ccaption{}{ \label{fig:pinch} 
The master integral topologies of the pinchings coming from the reduction of
the scalar crossed two-loop box.}
\end{figure}

At the end of this reduction program, we are left with the following two
crossed-box integrals: $\xbox^D(1,1,1,1,1,1,1;s,t)$ and
$\xbox^D(1,1,1,1,1,1,2;s,t)$, plus simpler diagrams that can
always be expressed as a combination of:
\begin{itemize}
\item[-] the master crossed triangle of
  Fig.~\ref{fig:pinch} (a)
  \beq
     \xtm{D}{s} = \xbox^D\(1,0,1,1,1,1,1;s,t\),
  \eeq
\item[-] the master diagonal box of
  Fig.~\ref{fig:pinch} (b), produced  by
  \beq
   \dbm{D}{s}{t} = \xbox^D\(0,1,1,0,1,1,1;s,t\)=\xbox^D\(1,1,0,1,1,1,0;s,t\) ,
  \eeq
  together with 
  \beq
   \dbm{D}{s}{u} = \xbox^D\(1,1,0,1,0,1,1;s,t\)=\xbox^D\(0,1,1,1,1,0,1;s,t\) ,
  \eeq
  and 
  \beq
   \dbm{D}{t}{u} = \xbox^D\(0,1,0,1,1,1,1;s,t\),
  \eeq
\item[-] the master box with a bubble insertion of
  Fig.~\ref{fig:pinch} (c), produced  by
  \beq
     \bbm{D}{s}{t} = \xbox^D\(1,1,1,0,1,1,0;s,t\),
  \eeq
   together with 
  \beq
     \bbm{D}{s}{u} = \xbox^D\(1,1,1,1,0,0,1;s,t\),
  \eeq
\item[-] the master triangle with a bubble insertion of
  Fig.~\ref{fig:pinch} (d), produced by
  \beq
     \tm1{D}{s} = \xbox^D\(1,0,1,0,1,1,0;s,t\)= \xbox^D\(1,0,1,1,0,0,1;s,t\),
  \eeq
\item[-] the master sunset diagram of
  Fig.~\ref{fig:pinch} (e), produced by
  \beq
     \ssm{D}{s} = \xbox^D\(0,0,1,0,1,0,1;s,t\)=\xbox^D\(1,0,0,1,0,1,0;s,t\),
  \eeq
  together with 
  \beq
     \ssm{D}{t} = \xbox^D\(0,1,0,0,1,1,0;s,t\),
  \eeq
  and
  \beq
     \ssm{D}{u} = \xbox^D\(0,1,0,1,0,0,1;s,t\),
  \eeq
\end{itemize}
where we explicitly used the symmetry of the crossed box 
\beq
\Xbox = \xbox^D\(\nu_3,\nu_2,\nu_1,\nu_7,\nu_6,\nu_5,\nu_4;s,t\).
\eeq
The reduction and dimensional shift of the generic scalar diagonal box and of
the box with a bubble insertion have been treated in Ref.~\cite{pentabox}.
The tensor reduction and dimensional shift of the crossed triangle is treated
in Appendix~\ref{app:Xtri}.  All the diagrams obtained from the pinchings of
the other propagators can be easily reduced to the previous ones using the
\IBP\ identities for the crossed box and for the crossed triangle, with the
corresponding powers of propagators set to zero.

Instead of keeping $\xbox^D(1,1,1,1,1,1,2;s,t)$ as one of the two members of
the basis, we prefer to switch to a more symmetric integral:
$\xbox^D(1,2,1,1,1,1,1;s,t)$. The expression for this one, in terms of the
other two, can be easily obtained through the application of the reduction
formalism outlined above
\beqn
\label{eq:xbm2_to_xbm4}
&&\xbox^D(1,2,1,1,1,1,1;s,t) =  
 c_1 \,\xbox^D(1,1,1,1,1,1,1;s,t)\nonumber
\\&&\hspace{1.5cm}{}
+c_2 \, \xbox^D(1,1,1,1,1,1,2;s,t)
+c_3 \, \xtm{D}{s}
+c_4 \, \dbm{D}{s}{t}
\nonumber
\\&&\hspace{1.5cm}{}
+c_5 \, \dbm{D}{s}{u}
+c_6 \, \dbm{D}{t}{u}
+c_7 \, \bbm{D}{s}{t}
+c_8 \, \bbm{D}{s}{u}
\nonumber
\\&&\hspace{1.5cm}{}
+c_9 \, \tm1{D}{s}
+c_{10} \, \ssm{D}{s}
+c_{11} \, \ssm{D}{t}
+c_{12} \, \ssm{D}{u},
\eeqn
where the coefficients $c_j$ are collected in Appendix~\ref{app:xbm2}. This
allows us to define as master integrals
\beqn
 \xbm1{D}{s}{t} &=& \xbox^D(1,1,1,1,1,1,1;s,t)\\
 \xbm2{D}{s}{t} &=& \xbox^D(1,2,1,1,1,1,1;s,t),
\eeqn
which are symmetric under the exchange $t \leftrightarrow u$. This will
produce more compact results in the rest of the paper.

\subsection{The dimensional shift for the two master integrals}
\label{sec:dim_shift}
Exploiting the equivalence between $\dminus$ and $\P$ of
Eq.~(\ref{eq:P_and_dm}), we can rewrite Eq.~(\ref{eq:pm_d})
using~(\ref{eq:P})  
\beqn
&&\xbox^{D}\(\nu_1,\nu_2,\nu_3,\nu_4,\nu_5,\nu_6,\nu_7;s,t\)
\nonumber\\
&&\hspace{2cm} {} = \Big[ \(\p7+\p6+\p5+\p4\) \(\p3+\p2+\p1\)
\nonumber\\
&&\hspace{2cm} {} 
+\(\p5+\p4\)\(\p7+\p6\) \Big]
            \xbox^{D+2}\(\nu_1,\nu_2,\nu_3,\nu_4,\nu_5,\nu_6,\nu_7;s,t\),
\nonumber\\
\eeqn
and more specifically
\beqn
\xbm1{D}{s}{t} &=& \Big[ \(\ps7+\ps6+\ps5+\ps4\) \(\ps3+\ps2+\ps1\)
\nonumber\\
&&{}  +\(\ps5+\ps4\)\(\ps7+\ps6\) \Big]
            \xbm1{D+2}{s}{t}, \\
\xbm2{D}{s}{t} &=& \Big[ \(\ps7+\ps6+\ps5+\ps4\) \(\ps3+2\,\ps2+\ps1\)
\nonumber\\
&&{}  +\(\ps5+\ps4\)\(\ps7+\ps6\) \Big]
            \xbm2{D+2}{s}{t}. 
\eeqn
The right-hand side of the two equations can then be reduced to a linear
combination of master integrals in $D+2$ dimensions, and the system can then
be inverted to give $\xbm1{D+2}{s}{t}$ and $\xbm2{D+2}{s}{t}$ as a function of
master integrals in $D$ dimensions    
\beqn
\label{eq:dimen1}
\xbm1{D+2}{s}{t} &=& A(t,u) + A(u,t),\\
\xbm2{D+2}{s}{t} &=& B(t,u) + B(u,t),
\eeqn
where
\beqn
A(t,u) &=&
 a_{1} \,\xbm1{D}{s}{t} +a_{2}\, \xbm2{D}{s}{t} +a_{3} \,\xtm{D}{s}
+a_{4} \,\dbm{D}{s}{t}
\nonumber \\ 
  &&{}  
+a_{5} \,\dbm{D}{t}{u} 
+a_{6} \,\bbm{D}{s}{t}
+a_{7} \,\tm1{D}{s}   
\nonumber \\ 
&&{}  
+a_{8} \,\ssm{D}{s}
+a_{9} \,\ssm{D}{t}, \nonumber\\
\label{eq:dimen2}
B(t,u)  &=& 
b_{1} \,\xbm1{D}{s}{t} 
+b_{2} \,\xbm2{D}{s}{t} 
+b_{3} \,\xtm{D}{s}  
+b_{4} \,\dbm{D}{s}{t}\nonumber \\ 
  &&{} 
+b_{5} \,\dbm{D}{t}{u} 
+b_{6} \,\bbm{D}{s}{t}
+b_{7} \,\tm1{D}{s} 
\nonumber \\ 
&&{}  
+b_{8} \,\ssm{D}{s}
+b_{9} \,\ssm{D}{t}.
\end{eqnarray}
Some of the coefficients $a_{i}$ and $b_{i}$ of the system are collected in
Appendix~\ref{app:dim_shift}.

\section{Differential equations for the two master integrals}
\label{sec:system}
In the previous sections we showed how tensor integrals can be expressed in
terms of the two master crossed boxes, $\xbox_1^D$ and $\xbox_2^D$ plus
simpler master diagrams. The analytic expansion in $\ep=(4-D)/2$ of the first
master integral was computed in Ref.~\cite{Bas}.
We can obtain the analytic form for the second one by writing the derivative
of $\xbox_1^D$ with respect to one of the two independent physical scales
(that we choose to be $t$), as a combination of master
integrals, and solving the equation for $\xbox_2^D$.

Moreover we can verify the correctness of both the expressions of
$\xbm1{D}{s}{t}$ and $\xbm2{D}{s}{t}$, by deriving an analogous differential
equation for $\xbox_2^D$, and checking that the obtained identity is satisfied.
 
Using the procedure outlined in Section~\ref{sec:tensor_redux}, 
we differentiate with respect to $t$ the two master integrals written in the
form of Eq.~(\ref{eq:Schwinger}).  In this way, the only $t$-dependence comes
from $\Q$ of Eq.~(\ref{eq:def_I}) and the result is
\beqn
\frac{\partial}{\partial t} \xbm1{D}{s}{t} &=& 
\xbox^{D+2}\(1,2,1,2,1,1,2;s,t\)-\xbox^{D+2}\(1,2,1,1,2,2,1;s,t\),\nonumber\\
\frac{\partial}{\partial t} \xbm2{D}{s}{t} &=& 
2\, \xbox^{D+2}\(1,3,1,2,1,1,2;s,t\)- 2\,\xbox^{D+2}\(1,3,1,1,2,2,1;s,t\).
\nonumber
\eeqn
Applying the reduction formulae of Section~\ref{sec:reduction} and the
dimensional-shift of Section~\ref{sec:dim_shift} to the right-hand sides of
the system, we can rewrite them as a combination of master crossed boxes and
pinchings in $D$ dimensions.

The final system of differential equations  in arbitrary $D$ is given by
\beqn
\label{eq:der_t_xbm1}
\frac{\partial}{\partial t} \xbm1{D}{s}{t} &=&
\frac{1}{t-u} \lq H(t,u) + H(u,t), \rq 
\\
\label{eq:der_t_xbm2}
\frac{\partial }{\partial t} \xbm2{D}{s}{t}  &=&
\frac{1}{t-u} \lq K(t,u) + K(u,t), \rq 
\eeqn
where
\beqn
H(t,u) &=&
 h_1\, \xbm1{D}{s}{t}
+h_2\, \xbm2{D}{s}{t}
+h_3\, \xtm{D}{s}
+h_4\, \dbm{D}{s}{t}
\nonumber \\
   && {}
+h_5\, \dbm{D}{t}{u}
+h_6\, \bbm{D}{s}{t}
+h_7\, \tm1{D}{s}
\nonumber \\
   && {}
+h_{8}\, \ssm{D}{s}
+h_{9}\, \ssm{D}{t},
\\
K(t,u) &=& 
 k_1\, \xbm1{D}{s}{t}
+k_2\, \xbm2{D}{s}{t}
+k_3\, \xtm{D}{s}
+k_4\, \dbm{D}{s}{t}
\nonumber \\
   && {}
+k_5\, \dbm{D}{t}{u}
+k_6\, \bbm{D}{s}{t}
+k_7\, \tm1{D}{s}
\nonumber \\
   && {}
+k_8\, \ssm{D}{s}
+k_{9}\, \ssm{D}{t},
\eeqn 
and the coefficients $h_i$ and $k_i$ are listed in
Appendix~\ref{app:diff_eqs}.

From the symmetry $t \leftrightarrow u$ of the two master integrals, we
expect the two derivatives to be anti-symmetric with respect to the exchange
$t\leftrightarrow u$, and this is a further check of the correctness of the
system.

\section{The off-shell method}
\label{sec:diff_eqs}
An alternative way to derive the system of equations~(\ref{eq:der_t_xbm1})
and~(\ref{eq:der_t_xbm2}) is based on the construction of a set of
differential equations for the crossed-box integrals with the momentum
$p_1$ taken off-shell.  This is done using the algorithms and computer
programs described in~\cite{GR}, by means of which the differential equations
for any massless two-loop four-point function with three light-like and one
off-shell leg can be obtained. We then take the on-shell limit $p_1^2\to 0$.

To derive the differential equations for the diagrams with
one leg off-shell, we take $p_2$, $p_3$ and $p_4$,
all massless and on-shell, as independent momenta,
and form the derivatives with respect to the external invariants
$s_{23}$, $s_{24}$, $s_{34}$, where $s_{ij}=p_{ij}^2$
\begin{eqnarray}
s_{23} \frac{\partial}{\partial s_{23}} & = & \frac{1}{2} \left( +
p_2^{\mu} \,\frac{\partial}{\partial p_2^{\mu}} +
p_3^{\mu} \,\frac{\partial}{\partial p_3^{\mu}} -
p_4^{\mu} \,\frac{\partial}{\partial p_4^{\mu}}\right), \nonumber \\
s_{24} \frac{\partial}{\partial s_{24}} & = & \frac{1}{2} \left( +
p_2^{\mu} \,\frac{\partial}{\partial p_2^{\mu}} -
p_3^{\mu} \,\frac{\partial}{\partial p_3^{\mu}} +
p_4^{\mu} \,\frac{\partial}{\partial p_4^{\mu}}\right),
\label{eq:derivatives} \\
s_{34} \frac{\partial}{\partial s_{34}} & = & \frac{1}{2} \left( - 
p_2^{\mu} \,\frac{\partial}{\partial p_2^{\mu}} +
p_3^{\mu} \,\frac{\partial}{\partial p_3^{\mu}} +
p_4^{\mu} \,\frac{\partial}{\partial p_4^{\mu}}\right) \; .\nonumber 
\end{eqnarray}
The derivatives with respect to $p_i^{\mu}$ act on the propagators in the
Feynman integrand of Eq.~(\ref{eq:Xbox_def}) and give rise to a number of
different integrals
with increased powers of the propagators. Using \IBP\ 
and Lorentz-invariance identities, all these different integrals
can be reduced to a small number of master integrals. In particular,
any crossed-box topology integral
can be reduced to a linear combination of two master
crossed-box integrals, which we can choose to be the off-shell versions
of $\xbox_1^{D}$ and $\xbox_2^{D}$, plus a number of pinched master
integrals. Thus, we obtain a system of two coupled linear first-order
differential equations for the two master crossed-box integrals.

Since, in the on-shell limit, we are interested in the derivatives
at fixed $p_1^2=0$, we rewrite the system of equations
in terms of the  variables
$p_1^2 = s_{23}+s_{24}+s_{34}$, $s=s_{34}$, $t=s_{23}$,
and the corresponding derivative operators:
\begin{eqnarray}
\frac{\partial}{\partial s} & = & \frac{\partial}{\partial s_{34}} -
                                  \frac{\partial}{\partial s_{24}},
\nonumber\\
\frac{\partial}{\partial t} & = & \frac{\partial}{\partial s_{23}} -
                                  \frac{\partial}{\partial s_{24}},
\label{eq:newder}\\
\frac{\partial}{\partial p_1^2} & = & \frac{\partial}{\partial s_{24}}\;.
\nonumber
\end{eqnarray}
The next step is to take the on-shell limit $p_1^2\to 0$. A complication
here is the fact that factors of $1/p_1^2$ and $1/(p_1^2)^2$ appear in
some of the coefficients of the pinched master integrals. It is also
important to realize that several of the off-shell master integrals with
less than seven propagators become reducible in the on-shell limit (an
example of such a case will be given below). When these reductions are
taken into account, all the terms proportional to
$1/p_1^2$ and $1/(p_1^2)^2$ cancel out, so that the on-shell
limit of the system of equations for the two master crossed-box integrals
is indeed well defined.


However, because of these factors of
$p_1^2$ in the denominator, it is not sufficient merely to replace all
pinched master integrals by their limit $p_1^2\to 0$: subleading terms
in $p_1^2$ also have to be included.
The Taylor series around $p_1^2=0$ for the pinched diagrams can be
obtained in a straightforward manner by using the differential equation in
$p_1^2$. As an example, we consider the propagator diagram
$\ssm{D}{p_1^2-s-t}$, which fulfills the homogeneous differential equation
\begin{equation}
\frac{\partial}{\partial p_1^2} \,\ssm{D}{p_1^2-s-t} = \frac{D-3}{p_1^2-s-t}
\,\ssm{D}{p_1^2-s-t}.
\end{equation}
Iterating this equation, we obtain 
\begin{eqnarray}
\ssm{D}{p_1^2-s-t} & = &
\ssm{D}{-s-t} + p_1^2 \, \frac{D-3}{-s-t}\, \ssm{D}{-s-t}
\nonumber\\ &&
              {}  + \frac{1}{2} \,p_1^4 \, \frac{(D-3)(D-4)}{(-s-t)^2}
\, \ssm{D}{-s-t} + {\cal O}\(p_1^6\) .
\end{eqnarray}

If a factor of $1/p_1^2$ appears in the homogeneous term
of the differential equations for an off-shell master integral, 
this integral becomes reducible in the limit $p_1^2\to 0$.
\begin{figure}[t]
\begin{center}
\epsfig{file=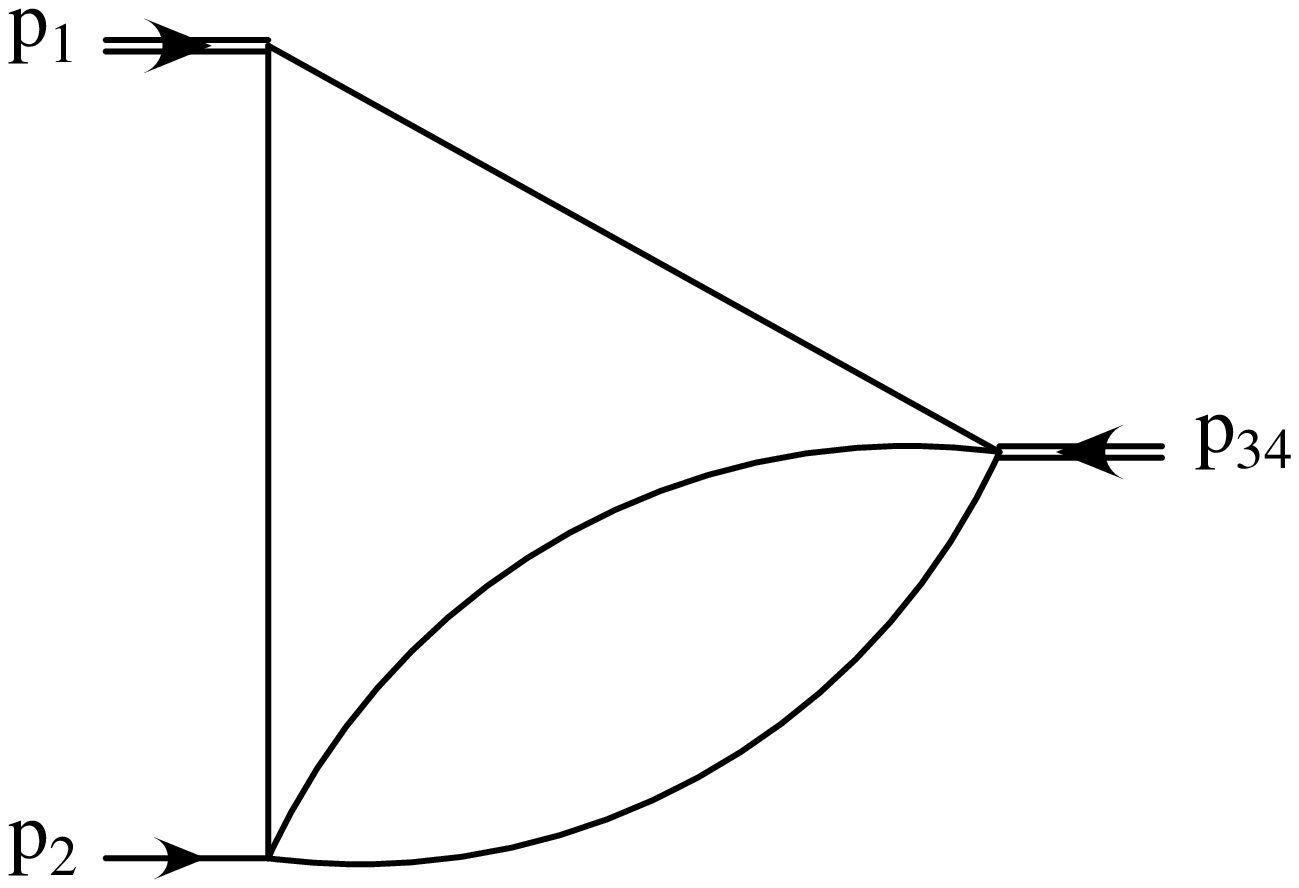,height=5cm}
\end{center}
\ccaption{}{\label{fig:offshellvtx}
The master integral $\rtri\(p_1^2,s\)$, which becomes reducible when
$p_1$ is on-shell.}
\end{figure}
An example of a 
reducible master integral is the vertex diagram
$\rtri\(p_1^2,s\)$ of Fig.~\ref{fig:offshellvtx}, which fulfils the
differential equation
\begin{equation}
\frac{\partial}{\partial p_1^2} \, \rtri\(p_1^2,s\) = \frac{D-4}{2} \,
\frac{2p_1^2-s}{p_1^2\,(p_1^2-s)} \, \rtri\(p_1^2,s\) - \frac{3D-8}{2} \,
\frac{1}{p_1^2\,(p_1^2-s)} \,\ssm{D}{s} .
\label{eq:diffBtri}
\end{equation}
Multiplying by $p_1^2$, one can immediately read off the limit 
\begin{equation}
 \rtri\(p_1^2\to 0,s\) = -\frac{3D-8}{D-4} \, \frac{1}{s}\, \ssm{D}{s} 
+  {\cal O}\(p_1^2\).
\end{equation}
Subleading terms in $p_1^2$ are again
obtained  by successive iteration of the differential equation~(\ref{eq:diffBtri}).

All the information required to derive the massless
limit of the system of equations for the master crossed-box
diagrams is contained in the differential equations for
the pinched master integrals.

The differential equations obtained with this method are in perfect agreement
with~(\ref{eq:der_t_xbm1}) and~(\ref{eq:der_t_xbm2}), giving an
independent confirmation of the correctness of the two methods.

In addition, we checked that the two methods give the same results for the
crossed boxes with the powers of propagators equal to unity and with one
scalar product in the numerator, of the type $p_i \cdot k$ or $p_i \cdot
l$. This was done computing Eq.(\ref{eq:Xbox_kmu}) with the method described
in Section~\ref{sec:tensor_redux} and then contracting the result with one of
the external momenta.

\section{Analytic expansion of the second master integral}
\label{sec:an_expans}
Inserting the $\ep$ expansion of $\xbm1{D}{s}{t}$ computed in Ref.~\cite{Bas}
and the $\ep$ expansions of the sub-topologies listed in
Refs.~\cite{Smirnov_Veretin,pentabox,Gonsalves83,Kramer} into
Eq.~(\ref{eq:der_t_xbm1}), and solving it with respect to $\xbm2{D}{s}{t}$,
we obtain, in the physical region $s>0$, $t,u <0$,
\begin{equation}
\label{eq:master2g1g2}
\xbm{2}{D}{s}{t} = \Gamma^2(1+\epsilon)
 \left\{
 \frac{G_1(t,u)}{s^3 t} +
 \frac{G_2(t,u)}{s^2 t^2} +
 \frac{G_1(u,t)}{s^3 u} +
 \frac{G_2(u,t)}{s^2 u^2}
 \right\}
\, ,
\end{equation}
where
\begin{eqnarray}
\label{eq:g1tu}
&& G_1(t,u) = s^{-2\epsilon} \left\{ \vphantom{\frac{1}{1}} \right.
\frac{6}{{\epsilon}^3}
 + \frac{1}{{\epsilon}^2} \left( 32 - 6\,T - 6\,U \right)
\nonumber \\ && \hspace{1cm}
 {}+ \frac{1}{\epsilon}
\left( 1 - 12\,{\pi }^2 - 24\,T + T^2 - 24\,U + 16\,T\,U + U^2 \right)
  -43 - 18\,T + 13\,T^2 + \frac{8}{3}\,T^3
\nonumber \\ && \hspace{1cm}
 {}   - 18\,U + 16\,T\,U + 11\,T^2\,U + 13\,U^2 - 20\,T\,U^2 + 
    \frac{8}{3}\,U^3
 + {\pi }^2
     \left( 17\,T + 17\,U  -\frac{112}{3} \right)
\nonumber \\ && \hspace{1cm}
   {}     - 122\,\zeta(3) + 62\,T\,\Li{2}{\mtos} - 
    62\,\Li{3}{\mtos}
 + 62\,\snp{1,2}{\mtos}
\nonumber \\ && \hspace{1cm}
{}+ i\pi \left[
\frac{1}{\epsilon} \left( 16 + 6\,T + 6\,U \right)
  -34 - 9\,{\pi }^2 - 6\,T - 10\,T^2 - 6\,U
    +  14\,T\,U - 10\,U^2
\left. \vphantom{\frac{1}{1}} \right] \right\}  ,
\nonumber \\ &&
\end{eqnarray}
\begin{eqnarray}
\label{eq:g2tu}
&& G_2(t,u) = s^{-2\epsilon} \left\{ \vphantom{\frac{1}{1}} \right.
-\frac{2}{{\epsilon}^4}
 + \frac{1}{{\epsilon}^3}
 \left(-8 + \frac{5}{2}\,T + 
     \frac{7}{2}\,U \right) 
\nonumber \\ && \hspace{1cm}
{} + \frac{1}{{\epsilon}^2}
 \left( -\frac{29}{2} - \frac{5}{12}\,{\pi }^2 + 
     7\,T - T^2 + 20\,U - 4\,T\,U - U^2 \right)
\nonumber \\ && \hspace{1cm}
{}  +  \frac{1}{\epsilon}
 \left[ -\frac{1}{2}  + 17\,T + 2\,T^2 - 
     \frac{T^3}{3} + \frac{{\pi }^2}{6}
      \left( 14 + 5\,T - 29\,U
        \right)  + 13\,U - 28\,T\,U - 4\,U^2
\right. \nonumber \\ && \hspace{1cm}
\hspace{1em} \left.
{} + 3\,T\,U^2 - U^3 + 
     \frac{19}{2}\,\zeta(3) - 2\,T\,\Li{2}{\mtos} + 
     2\,\Li{3}{\mtos} - 2\,\snp{1,2}{\mtos} \right]
\nonumber \\ && \hspace{1cm}
{} + \frac{37}{2} + \frac{37}{40}\,{\pi }^4 + 7\,T - 5\,T^2 - 
    \frac{22}{3}\,T^3 + \frac{2}{3}\,T^4 + 5\,U - 20\,T\,U + 
    \frac{8}{3}\,T^3\,U - 2\,U^2
\nonumber \\ && \hspace{1cm}
{} + 24\,T\,U^2 - T^2\,U^2 - 
    8\,U^3 - \frac{4}{3}\,T\,U^3 + \frac{4}{3}\,U^4
\nonumber \\ && \hspace{1cm}
 {}  +   \frac{{\pi }^2}{6}\left( 79 - 22\,T - 
       5\,T^2 - 200\,U + 76\,T\,U + 
       25\,U^2 \right)
  + \left( 68 - 13\,T - 33\,U \right) \,\zeta(3)
\nonumber \\ && \hspace{1cm}
{} + \left( 10\,{\pi }^2 - 32\,T + 17\,T^2 + 12\,T\,U \right) \,
     \Li{2}{\mtos}
 + \left( 32 - 60\,T - 12\,U \right) \,
     \Li{3}{\mtos}
\nonumber \\ && \hspace{1cm}
{} + \left(28\,T - 6\,U  -32 \right) \,\snp{1,2}{\mtos}
 - 26\,\snp{1,3}{\mtos} - 36\,\snp{2,2}{\mtos}
 + 86\,\Li{4}{\mtos}
\nonumber \\ && \hspace{1cm}
{}+ i\pi \left[
\frac{2}{{\epsilon}^3}
 + \frac{1}{{\epsilon}^2}\left( 11 - T + U \right)
 + \frac{1}{\epsilon}
 \left( 1 - \frac{31}{6} \,{\pi }^2 - 10\,T - 2\,T^2 + 4\,U - 
     2\,T\,U - 2\,U^2 \right)
\right. \nonumber \\ && \hspace{1cm}
\hspace{1em}
{} + 11 + 4\,T - 2\,T^2 + \frac{10}{3}\,T^3 + 
    \frac{{\pi }^2}{3} \left( - 65
  + 28\,T - U \right)  + 2\,U - 8\,T\,U - 
    8\,U^2
\nonumber \\ && \hspace{1cm}
\hspace{1em} \left.
{} + 2\,U^3 - 89\,\zeta(3) + 
    \left( 14\,T + 18\,U \right) \,\Li{2}{\mtos} - 
    32\,\Li{3}{\mtos} + 44\,\snp{1,2}{\mtos}
\left. \vphantom{\frac{1}{1}} \right] \right\}  ,
\nonumber \\ &&
\end{eqnarray}
and $T=\log(-t/s)$, $U=\log(-u/s)$.
We used  Nielsen's generalized polylogarithms
$\mbox{S}_{n,p}$~\cite{Nielsen,KolbigMignacoRemiddi70}, defined by
\begin{equation}
\snp{n,p}{x} = \frac{(-1)^{n+p-1}}{(n-1)!\,p!}
\int_0^1 \mbox{d} t \; \frac{ \log^{n-1}(t) \log^p(1-xt) }{t},
\quad\quad n,p \ge 1, \quad x \le 1,
\end{equation}
where the usual polylogarithms are given by
\beq
{\rm Li}_n(x) = \snp{n-1,1}{x}.
\eeq

\begin{figure}[t]
\begin{center}
\epsfig{file=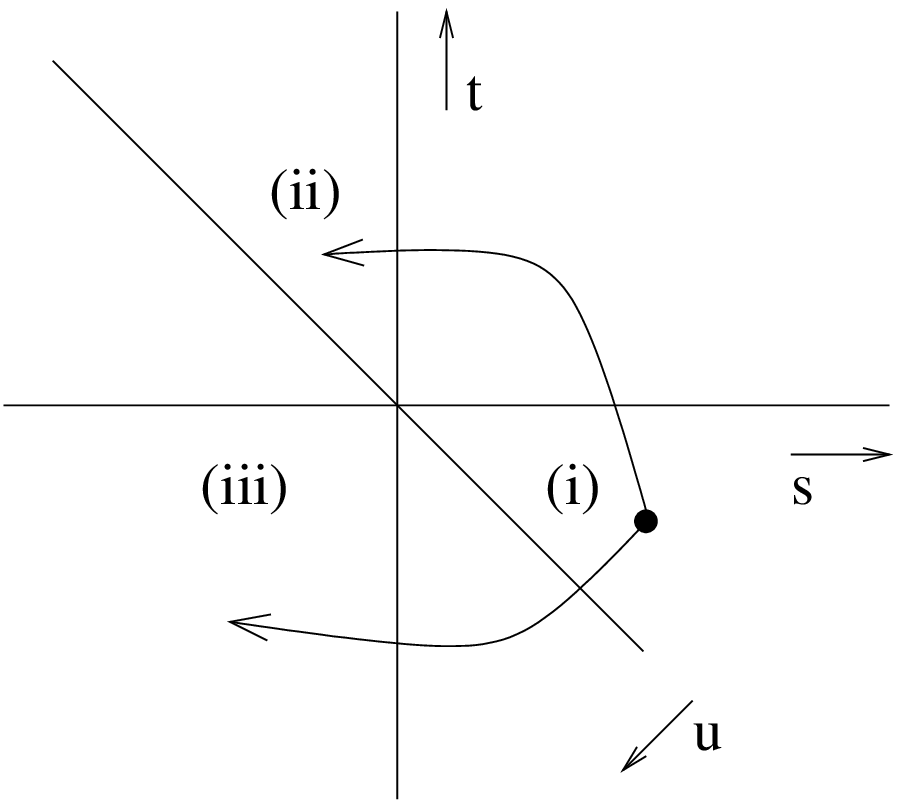,height=6cm}
\end{center}
\ccaption{}{\label{fig:regs123}
The physical regions {\rm (i), (ii)} and {\rm (iii)} in the $(s,t,u)$-plane.}
\end{figure}

The three kinematically accessible regions of the phase-space are depicted in
Fig.~\ref{fig:regs123}. 

\begin{itemize}
\item[\bf (i)] $\boldsymbol{s>0,\ t,u<0}$. All
 logarithms and polylogarithms occurring in Eqs.~(\ref{eq:g1tu})
 and~(\ref{eq:g2tu}) are real in this region.

 Formulae for the other two regions, (ii) and (iii), can be derived by
 analytic continuation, starting from region (i) and following the paths
 indicated in the figure.

The analytic continuation can be performed through a few simple steps.

\item[\bf (ii)] $\boldsymbol{t>0,\ s,u<0}$. 
 Going from region (i) to region (ii), we have to pass through two branches:
 $t=0$ and $s=0$. We can then split the analytic continuation into two steps:
\begin{itemize}
\item[-]
 we first split the logarithm $T= \log(-t)-\log(s)$. At $t=0$, nothing
 happens to the polylogarithms $\snp{n,p}{-t/s}$, but $\log(-t)$ gets an
 imaginary part: $\log(-t) \to \log(t) - i \pi$.

 We are now in an unphysical region, where both $s$ and $t$ are positive and
 $u$ is negative. Using the transformation formulae for $x \, \to\, 1/x$
 (see, eg. Refs.~\cite{KolbigMignacoRemiddi70,pentabox}), we can express
 $\snp{n,p}{-t/s}$ in terms of $\snp{n,p}{-s/t}$ and $\log(t/s)$.
 
\item[-] To enter region (ii), we have to pass now the branch point at $s=0$.
  We split $\log(t/s)=\log(t)-\log(s)$ and $U = \log(s+t)-\log(s)$ and
  we analytically continue $\log(s) \to \log(-s) + i \pi$.
\end{itemize}
In this way, for example, the logarithms in Eqs.~(\ref{eq:g1tu})
and~(\ref{eq:g2tu}) undergo the transformation
\beqn
 \log\(\mtos\) &\to&  \log\(-\frac{t}{s}\) -2\,i\,\pi,\\
\log\(-\frac{u}{s}\)  &\to & \log\(\frac{u}{s}\) -i\,\pi.
\eeqn

\item[\bf (iii)] $\boldsymbol{u>0,\ s,t<0}$. 
The procedure to go from region (i) to region (iii) is similar to the
previous one, but it requires an additional step. 
\begin{itemize}
\item[-] 
 We  rewrite  $\snp{n,p}{-t/s}$ in terms of
 $\snp{n,p}{(s+t)/s}$, $\log(-t/s)$ and $\log((s+t)/s)$, using the
 transformation $x \,\to \, 1-x$, and we split the logarithms as before.  In
 passing the first branch point at $u=0$, the polylogarithms are well
 defined while $\log(s+t) \to \log(-s-t) - i \pi$.
\item[-] 
 We invert now the argument of the polylogarithms, expressing
 $\snp{n,p}{(s+t)/s}$ in terms of $\snp{n,p}{s/(s+t)}$ and
 $\log((-s-t)/s)=\log(-s-t)-\log(s)$.  Finally, $\log(s) \to \log(-s) + i
 \pi$, as we pass the branch point at $s=0$ and enter region (iii).
\end{itemize}
The logarithms in Eqs.~(\ref{eq:g1tu}) and~(\ref{eq:g2tu}) undergo the
transformation
\beqn
 \log\(\mtos\) &\to&  \log\(\frac{t}{s}\) - i\,\pi,\\
\log\(-\frac{u}{s}\)  &\to & \log\(-\frac{u}{s}\) -2\,i\,\pi.
\eeqn
 
The expression for $G_1(t,u)$ and $G_2(t,u)$ in this region can also be
obtained directly from the expressions in region (ii), using the symmetry $t
\leftrightarrow u$.
\end{itemize}

A non-trivial check of the correctness of  the expressions of
$\xbm1{D}{s}{t}$ and $\xbm2{D}{s}{t}$ comes from Eq.~(\ref{eq:der_t_xbm2}),
that must be identically satisfied, once the respective $\ep$ expansions are
used.


\section{Conclusions}
\label{sec:conclusions}
High-energy scattering processes are one of the most important sources of
information on short-distance physics.
Recent improvements of experimental measurements demand nowadays the
knowledge of $2 \,\to\, 2$ scattering rates at two-loop order.

Two non-trivial topologies characterize these processes: planar two-loop
boxes~\cite{Smirnov,Smirnov_Veretin}, and crossed two-loop boxes.

The crossed boxes constituted the last barrier towards the completion of this
goal.  In this paper we presented an algorithm to reduce tensor two-loop
massless crossed boxes with light-like external legs to two master crossed
boxes (the first one with all powers of propagators equal to one, and the
second one with the power of the second propagator of Fig.~\ref{fig:Xboxfig}
equal to two), plus simpler diagrams.

We derived the equations that connect the two master integrals in dimensions
$D$ and $D+2$ and the system of first-order coupled differential equations
satisfied by the two master integrals.

This last part was done following two different methods: using the raising
and lowering operators, and using the on-shell limit of the differential
equations for the crossed box with one off-shell leg~\cite{GR}.  The
agreement between the results of the two methods is a strong support for the
validity of the two different procedures.

The differential equation for the first master integral allowed us to derive
an $\ep$ expansion for the second master integral, once the known~\cite{Bas}
expression of the first master integral is inserted.
The differential equation for the second master integral then provided us a
non-trivial way to check the obtained expansion.


\subsection*{Acknowledgement}

We thank E.W.N.~Glover, M.E.~Tejeda-Yeomans and J.J.~van der Bij for assistance
and useful suggestions.  C.A. acknowledges the financial support of the Greek
Government, C.O. acknowledges the financial support of the INFN, E.R. thanks
the Alexander-von-Humboldt Stiftung for supporting his stay at the Institut
f\"ur Theoretische Teilchenphysik of the University of Karlsruhe and
J.B.T. acknowledges the financial support of the DFG-Forschergruppe
``Quantenfeldtheorie, Computeralgebra und Monte-Carlo-Simulation''.  We
gratefully acknowledge the support of the British Council and German Academic
Exchange Service under ARC project 1050.

\appendix

\section{Tensor reduction and dimensional shift of the crossed two-loop
triangle} 
\label{app:Xtri}
\begin{figure}[htb]
\centerline{\epsfig{figure=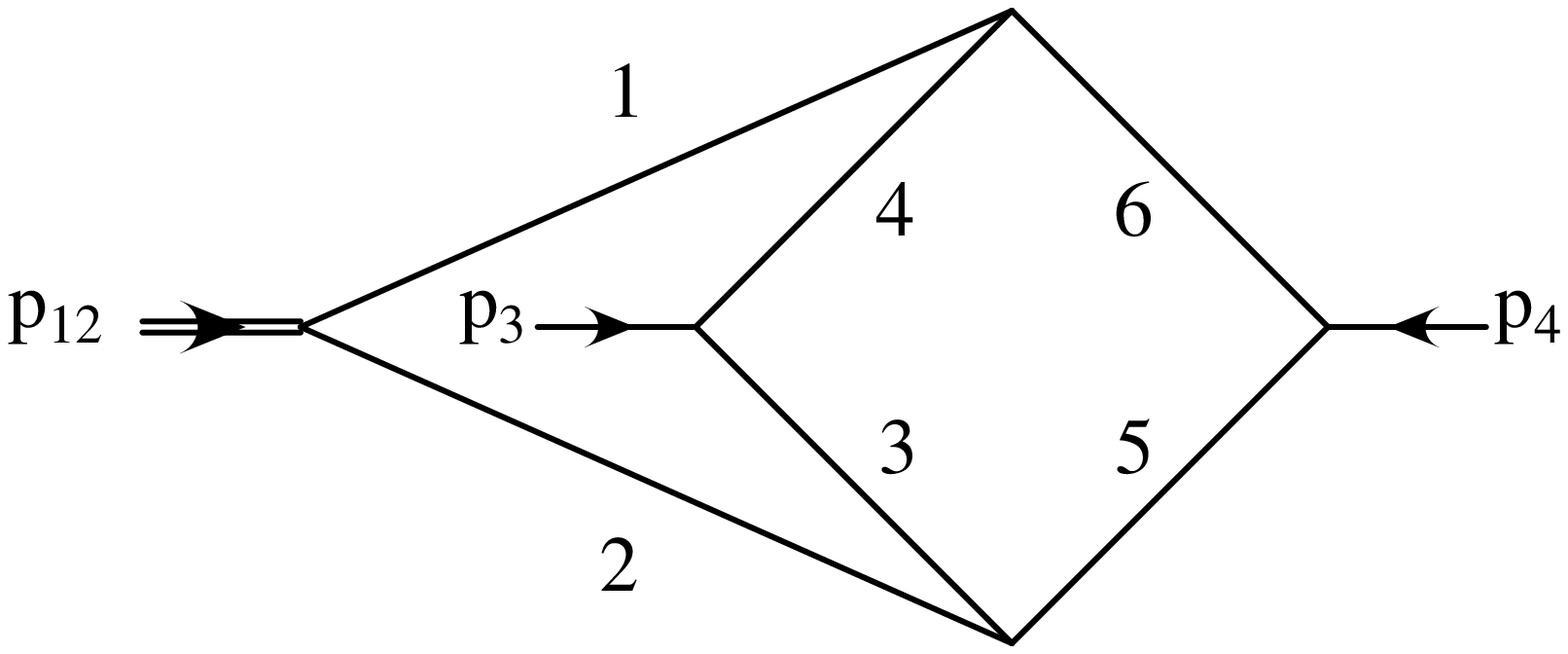,width=0.65\textwidth,clip=}}
\ccaption{}
{ \label{fig:Xtrifig} The generic two-loop crossed  triangle. }
\end{figure}

For completeness, we give here the reduction formulae for the crossed
two-loop triangle that appears as a sub-topology in the scalar reduction of
the crossed box. We make use of \IBP\ and Lorentz-invariance identities to
reduce the generic scalar triangle to one master integral which has all
powers of propagators equal to one~\cite{Gonsalves83,Kramer}, plus pinchings.
For an alternative solution to this reduction problem, exploiting
a connection with massless three-loop propagator integrals, see
Ref.~\cite{Baikov_Smirnov}.

The generic two-loop scalar crossed triangle of
Fig.~\ref{fig:Xtrifig} is given by
\beq
\label{eq:Xtri}
\xtri^D\( \n{1},\n{2},\n{3},\n{4},\n{5},\n{6}; s \)  = 
\int \frac{d^D k}{i \pi^{D/2}}
\int \frac{d^D l}{i \pi^{D/2}}
~
\frac{1}{
A_1^{\nu_1}
A_2^{\nu_2}
A_3^{\nu_3}
A_4^{\nu_4}
A_5^{\nu_5}
A_6^{\nu_6}},
\end{equation}
where the propagators are
\begin{eqnarray}
\label{eq:props}
A_1 &=& (k+l+p_3+p_4)^2 + i0,\\
A_2 &=& (k+l)^2 + i0, \nonumber \\
A_3 &=& l^2 + i0, \nonumber \\
A_4 &=& (l+p_3)^2 + i0, \nonumber \\
A_5 &=& k^2 + i0, \nonumber \\
A_6 &=& (k+p_4)^2 + i0. \nonumber 
\end{eqnarray}
The external momenta are in-going, two of them being light-like,
$p_3^2=p_4^2=0$, while $p_{12}^2=s$ is the only physical scale.
  
Repeating the reasoning of Section~\ref{sec:reduction}, we can build eight
\IBP\ identities and a single Lorentz-invariance identity. They depend upon one
irreducible scalar product in the numerator, that we choose to be $(l\cdot
p_4)$: 
\begin{eqnarray}
\label{eq:Ibpxtr1}
&& s \p{1} +\( 2 D -2 \n{235} -\n{146}\) -\p{4} \m{3} -\p{1}
\m{2}  - \p{6} \m{5}=0    \\
\label{eq:Ibpxtr2}
&& s \p{2} + \( 2 D -2 \n{146} -\n{235}\) -\p{3} \m{4} -\p{2}
\m{1} -\p{5} \m{6}=0  \\
&&  2\( l \cdot p_4\) \p{1}  - \(D-\n{24}-2 \n{3}\)  + 
\p{1} \(\m{2} +\m{4} -\m{5} \) 
 \nonumber \\
&& \hspace{2cm} {} + \p{2} \(\m{3} -\m{5} \)  +\p{4} \m{3}=0 \\
&& 2\( l \cdot p_4\) \p{2}  - \(D-\n{125}-2 \n{6} \)  + \p{2} \(\m{6} -\m{3} \)  \nonumber \\
&& \hspace{2cm} {} -\p{1} \(\m{6} - \m{4} \)  - \p{5} \m{6} =0 \\
&& 2\( l \cdot p_4\) \p{3}  + \(D-\n{345}-\n{6}\)  + \p{3} \(\m{2}-\m{5} \) 
+\p{4} \(\m{1} - \m{6}\) \nonumber \\
&&\hspace{2cm} {}  -\p{5} \m{6}  =0 \\ 
&& 2\( l \cdot p_4\) \p{4}  + s \p{4}   -(D-\n{346}-2\n{5})  
+ \p{3} \( \m{5} -\m{2}\)
 \nonumber \\
&&\hspace{2cm} {} + \p{4} \(\m{6} -\m{1} \)  + \p{6}\m{5}=0 \\
&& 2\( l \cdot p_4\) \p{5}  +s \p{5}  - \(D-\n{36} -2 \n{4} \)  + \p{5}
 \(\m{4}+\m{6}  -\m{1} \) 
\nonumber \\
&& \hspace{2cm} {} +\p{6} \(\m{4} -\m{1} \)  + \p{3} \m{4}  =0  \\
&&  2\( l \cdot p_4\) \p{6} -  \(D-\n{45}-2 \n{3} \) - 
\p{6} \(\m{3} + \m{5} -\m{2} \) 
 \nonumber \\
&& \hspace{2cm} {}  +\p{5} \(\m{2} -\m{3} \)  + \p{4} \m{3} =0 \\
&& 2\( l \cdot p_4\) \p{1}  -\( D-\n{2356} \)  + \p{1} \( \m{2} +\m{6}
-\m{5}\)   + \p{4} \m{3}  =0.  
\end{eqnarray}
By eliminating the irreducible scalar product in the numerator, we obtain
\beq
s \p{1} = -\( 2 D -2 \n{235} -\n{146}\) +\p{4} \m{3} +\p{1}\m{2}+\p{6} \m{5},
\eeq
which, together with the symmetric one for $\p{2}$, can reduce $\n{1}$ and
$\n{2}$ to unity. 
Using
\beqn
\label{eq:Ibpxtr3}
\p{3}  &=& \frac{1}{D-2-\n{3456}} \(\p{4} \p{6} \m{1} - \p{3} \p{5}
  \m{2} \)  \nonumber \\
  && {}+\frac{1}{D-2-2 \n{34}} \lq \( D-2-2 \n{46}\) \p{6} +2 \(\n{3}-\n{6} \)
  \p{5} \rq,
\eeqn
and the symmetric one for $\p4$, we can reduce $\n{3}$ and $\n{4}$ to one.
To complete the reduction, we use
\beq
\label{eq:Ibpxtr5}
\( \n{56}-\n{34}\) = \p{2} \( \m{3}-\m{5} \)  + \p{1} \(\m{4}-\m{6}\),
\eeq
that can be re-iterated until $\( \n{56}-\n{34}\)=0$. Since we are applying
this identity to scalar integrals where $\n{3}$ and $\n{4}$ have already been
reduced to one, the reduction procedure will stop when
$\n{5}=\n{6}=1$.  This integral cannot be reduced any further, and we choose
the crossed master triangle to be
\beq
\label{eq:xtm}
\xtri^D(s) = \xtri^D\(1,1,1,1,1,1;s\).
\eeq
The application of the identity
\beq
\dminus = \(\p6+\p5+\p4+\p3\) \(\p2+\p1\)+\(\p4+\p3\)\(\p6+\p5\)
\eeq
to $\xtri^{D+2}(s)$ and its further reduction, gives rise to the
dimensional-shift formula
\beqn
\xtm{D+2}{s} &=& 
- \frac{(D-4)  s^2}{4 (D-2) (2 D-7) (2 D-5)} \,\xtm{D}{s}
\nonumber
\\&& {}	-\frac{37 D^3-313 D^2+858 D-752}
	 {2 (D-4) (D-2) (2 D-7) (2 D-5) (3 D-8)} \, \tm1{D}{s}
\nonumber
\\&& {}+ \frac{43 D^4-478 D^3+1963 D^2-3530 D+2352}
	 {2 (D-4)^2 (D-3) (D-2) (2 D-7) (2 D-5) s} \, \ssm{D}{s}.
\eeqn
The expression of the master integral of Eq.~(\ref{eq:xtm}) has been computed
in Refs.~\cite{Gonsalves83,Kramer}.

\section{Coefficients $\boldsymbol{c_i}$}
\label{app:xbm2}
We collect here the coefficients $c_i$ of Eq.~(\ref{eq:xbm2_to_xbm4}).
\beqn
c_1 &=&  4 \frac{(D-5)^2  u}{(D-6) s t} \nonumber
\\
c_2 &=& \frac{t-u}{t}\nonumber
\\
c_3 &=& -2 \frac{(D-4) (2 D-9)} {(D-6) s t}  \nonumber
\\
c_4 &=& -3 \frac{(D-4) (3 D-14) u^2}{(D-6)^2 (2 D-11) s^3 t^3}
	 \lq 2 \(7 D^2-76 D+206\) t+(D-5) (5 D-28) s\rq
\nonumber
\\
c_5 &=& -3 \frac{(D-4) (3 D-14)  t}{(D-6)^2 (2 D-11) s^3 u^2}
	 \lq 2 \(7 D^2-76 D+206\) t+(D-5) (5 D-28) s\rq
\nonumber
\\
c_6 &=& 3 \frac{(D-5) (D-4) (3 D-14)  s
	 \lq 2 (D-5) t-(D-6) s \rq }{
       (D-6)^2 (2 D-11) t^3 u^2}\nonumber
\\
c_7 &=& 6 \frac{ (D-3) (3 D-14)  \lq (5 D-26) t+2 (D-5) s \rq }
       {(D-6)^2 s t^3} \nonumber
\\
c_8 &=& 6 \frac{ (D-3) (3 D-14)  \lq (5 D-26) t+2 (D-5) s \rq}
       {(D-6)^2 s t u^2} \nonumber
\\
c_9 &=& -\frac{3}{2} \frac{(D-3) (3 D-10)} 
{(D-6)^2 (D-5) (D-4) s^3 t^3 u^2}
	 \Big[-\(41 D^3-620 D^2+3124 D-5248\) s t^3 \nonumber
\\ &&{}
+2 (D-6)^2 (D-4) t^4
			     -\(61 D^3-922 D^2+4640 D-7776\) s^2 t^2\nonumber
\\ &&{}
			     -4 (D-5) (3 D-14) (4 D-21) s^3 t
			     -4 (D-5)^2 (3 D-14) s^4 \Big]
       \nonumber
\\
c_{10} &=& -3 \frac{(D-3) (3 D-10) (3 D-8)}
{(D-6)^2 (D-5) (D-4)^2 (2 D-11) s^4 t^3 u^2} \nonumber
\\ &&{}
   \times     \Big[ 2 (D-6) (D-4) (2 D-11) (5 D-22) t^4\nonumber
\\ &&{}
      -2 \(64 D^4-1305 D^3+9978 D^2-33902 D+43180\) s t^3\nonumber
\\ &&{}
      -\(208 D^4-4209 D^3+31899 D^2-107314 D+135216\) s^2 t^2\nonumber
\\ &&{}
      -(D-5) (3 D-14) \(34 D^2-359 D+944\) s^3 t
      -(D-5)^2 (3 D-14) (5 D-28) s^4 \Big]
       \nonumber
\\
c_{11} &=& 
      3 \frac{(D-3) (3 D-14) (3 D-10) (3 D-8) }
 {(D-6)^2 (D-5) (D-4)^2 (2 D-11) s^3 t^4 u^2}\nonumber
\\ &&{}
	\times \Big[ 2 (D-5) \(7 D^2-76 D+206\) t^4
       +\(61 D^3-952 D^2+4941 D-8528\) s t^3\nonumber
\\ &&{}
       +\(92 D^3-1417 D^2+7253 D-12336\) s^2 t^2\nonumber
\\ &&{}
       +\(56 D^3-853 D^2+4317 D-7258\) s^3 t
       +2 (D-5) (2 D-11) (2 D-9) s^4 \Big]      
 \nonumber
\\
c_{12} &=& -3 \frac{(D-3) (3 D-14) (3 D-10) (3 D-8)}
{(D-6)^2 (D-5) (D-4)^2 (2 D-11) s^3 t^3 u^3}
	 \Big[ 2 (D-5) \(7 D^2-76 D+206\) t^4
\nonumber
\\ &&{}-\(9 D^3-130 D^2+615 D-948\) s t^3 -3 (D-6) (D-5) (D-4) s^2 t^2
\nonumber
\\ &&{}-(D-5) \(3 D^2-30 D+74\) s^3 t+(D-6) (D-5)^2 s^4 \Big]       
\eeqn

\section{Coefficients of the dimensional-shift system}
\label{app:dim_shift}
We give here only the first three coefficients of the
system~(\ref{eq:dimen2}). The whole list can be obtained from the
authors (C.A. and C.O.).

{
\beqn
a_{1} &=& \frac{3}{512}\frac{
 (D-4) s^2 \lq(3 D-14) (3 D-10) s^2  -4 (5 D^2-39 D+74) t u
			       \rq}
			{(D-3)^3 (2 D-9) (2 D-7) t u}
\nonumber\\
a_{2} &=& \frac{(D-6) (D-4) s\lq4 (D-3) t u -3 (3 D-10) s^2\rq}
			{512 (D-5) (D-3)^3 (2 D-9) (2 D-7)}
\nonumber\\
a_{3} &=& \frac{(D-4) s \lq 4 (D-6) (D-3) t u  + 3 (3 D-14) (3
	D-10) s^2\rq}{512 (D-5) (D-3)^3 (2 D-7) t u}
\nonumber\\
b_{1} &=& \frac{s \lq 3 (D-4) (3 D-14) s^2 - 4 (D-6) (D-5) t u \rq}
			{128 (D-3) (2 D-9) (2 D-7) t u}
\nonumber\\
b_{2} &=& -\frac{(D-6) \lq 4 (D-3)t u + 3 (D-4) s^2\rq}
			{128 (D-5) (D-3) (2 D-9) (2 D-7)}
\nonumber\\
b_{3} &=& \frac{3 (D-4) (3 D-14) s^2 - 4 (D-6) (D-3) t u}
			{128 (D-5) (D-3) (2 D-7) t u}
\eeqn
}

\section{Coefficients of the differential equations for the two master\\
integrals} 
The coefficients of the system of differential
equations~(\ref{eq:der_t_xbm1}) and~(\ref{eq:der_t_xbm2}) for the 
two master integrals are given by
\label{app:diff_eqs}
{
\beqn
h_1 &=& \frac{(D-4) s^2-4 t u}{4 t u} \nonumber
\\
h_2 &=& -\frac{(D-6) s}{4 (D-5)}\nonumber
\\
h_3 &=& \frac{(D-4) (2 D-9) s}{4 (D-5) t u}\nonumber
\\
h_4 &=& \frac{3}{2} \frac{(D-4) (3 D-14) u}{ (D-5) s t^2}\nonumber
\\
h_5 &=& \frac{3}{2} \frac{(D-4) (3 D-14) s^2}{ (D-6) t^2 u^2}\nonumber
\\
h_6 &=& 3 \frac{ (D-3) (3 D-14)}{(D-5) t^2}\nonumber
\\
h_7 &=& \frac{3}{4} 
\frac{(D-3) (3 D-10) \lq (3 D-14) \(u^2+t^2\)+2 (D-4) t u\rq}
 { (D-5) (D-4) s t^2 u^2}\nonumber
\\
h_8 &=& \frac{3}{4} \frac{ (D-3) (3 D-10) (3 D-8) 
\lq (D-5) (3 D-14) \(u^2+t^2\) -(D-6) (D-4) t u \rq }
 { (D-5)^2 (D-4)^2 s^2 t^2 u^2}\nonumber
\\
h_9 &=& \frac{3}{2} \frac{(D-3) (3 D-14) (3 D-10) (3 D-8)}
 {(D-6) (D-5)^2 (D-4)^2 s t^3 u^2}  
 \Big[ (2 D-9) (3 D-16) u^2\nonumber\\
&&{} +\(7 D^2-68 D+164\) t u+2 (D-5)^2 t^2 \Big],
\eeqn
and by
\beqn
k_1 &=&  \frac{(D-5)^2 s}{ t u}\nonumber
\\
k_2 &=& -\frac{(D-6) (u^2+t^2)}{2 t u} \nonumber
\\
k_3 &=&  \frac{(D-4) (2 D-9)}{ t u} \nonumber
\\
k_4 &=& 6 \frac{(D-4) (3 D-14) u \lq (5-D) u+(2 D-11) t \rq}
	 {(D-6) s^2 t^3}\nonumber
\\
k_5 &=& \frac{3}{2} \frac{(D-5) (D-4) (3 D-14) s^3}{ (D-6) t^3 u^3}\nonumber
\\
k_6 &=& 3 \frac{(D-3) (3 D-14) }
	 {(D-6) s t^3 u} \lq(5 D-28) t u+(D-6) t^2 -2 (D-5) u^2 \rq\nonumber
\\
k_7 &=& \frac{3}{2} \frac{(D-3) (3 D-10)}
{ (D-6) (D-4) s^2 t^3 u^3}
 \Big[  2 (D-6) (3 D-14) t u\(u^2+t^2\)\nonumber
\\ &&{}
    -(D-5) (3 D-14) \(u^4+t^4\)
			     +2 \(5 D^2-49 D+118\) t^2 u^2 \Big]
 \nonumber
\\
k_8 &=& 3 \frac{(D-3) (3 D-10) (3 D-8) }
 { (D-6) (D-5) (D-4)^2 s^3 t^3 u^3}
  \Big[ 3 (D-5)^2 (3 D-14) t u \( u^2+t^2 \)\nonumber
\\ && {} -(D-5)^2 (3 D-14) \(u^4+t^4\)
			       -(D-4) \(7 D^2-70 D+176\) t^2 u^2 \Big] 
	\nonumber
\\
k_9 &=& 3 \frac{(D-3) (3 D-14) (3 D-10) (3 D-8)}
{(D-6)^2 (D-5) (D-4)^2 s^2 t^4 u^3 }
  \Big[ (D-5)^2 (D-2) u^4 \nonumber
\\&&{}
+(D-6) \(13 D^2-129 D+318\) t u^3
			     +2 \(5 D^3-80 D^2+422 D-734\) t^2 u^2\nonumber
\\&&{}
			     +(D-6) (D-5) (5 D-24) t^3 u
			     +(D-6) (D-5)^2 t^4 \Big].
\eeqn
}

\relax
\def\pl#1#2#3{{\it Phys.\ Lett.\ }{\bf #1}\ (#2)\ #3}
\def\zp#1#2#3{{\it Z.\ Phys.\ }{\bf #1}\ (#2)\ #3}
\def\prl#1#2#3{{\it Phys.\ Rev.\ Lett.\ }{\bf #1}\ (#2)\ #3}
\def\rmp#1#2#3{{\it Rev.\ Mod.\ Phys.\ }{\bf#1}\ (#2)\ #3}
\def\prep#1#2#3{{\it Phys.\ Rep.\ }{\bf #1}\ (#2)\ #3}
\def\pr#1#2#3{{\it Phys.\ Rev.\ }{\bf #1}\ (#2)\ #3}
\def\np#1#2#3{{\it Nucl.\ Phys.\ }{\bf #1}\ (#2)\ #3}
\def\sjnp#1#2#3{{\it Sov.\ J.\ Nucl.\ Phys.\ }{\bf #1}\ (#2)\ #3}
\def\app#1#2#3{{\it Acta Phys.\ Polon.\ }{\bf #1}\ (#2)\ #3}
\def\jmp#1#2#3{{\it J.\ Math.\ Phys.\ }{\bf #1}\ (#2)\ #3}
\def\jp#1#2#3{{\it J.\ Phys.\ }{\bf #1}\ (#2)\ #3}
\def\nc#1#2#3{{\it Nuovo Cim.\ }{\bf #1}\ (#2)\ #3}
\def\lnc#1#2#3{{\it Lett.\ Nuovo Cim.\ }{\bf #1}\ (#2)\ #3}
\def\ptp#1#2#3{{\it Progr. Theor. Phys.\ }{\bf #1}\ (#2)\ #3}
\def\tmf#1#2#3{{\it Teor.\ Mat.\ Fiz.\ }{\bf #1}\ (#2)\ #3}
\def\tmp#1#2#3{{\it Theor.\ Math.\ Phys.\ }{\bf #1}\ (#2)\ #3}
\def\jhep#1#2#3{{\it J.\ High\ Energy\ Phys.\ }{\bf #1}\ (#2)\ #3}
\def\epj#1#2#3{{\it Eur.\ Phys. J.\ }{\bf #1}\ (#2)\ #3}
\relax

\end{document}